\documentclass[12pt,preprint]{aastex}
\usepackage{natbib}
\newcommand\lsim{\mathrel{\rlap{\lower4pt\hbox{\hskip1pt$\sim$}}
        \raise1pt\hbox{$<$}}}
\newcommand\gsim{\mathrel{\rlap{\lower4pt\hbox{\hskip1pt$\sim$}}
        \raise1pt\hbox{$>$}}}

\newcommand{\K}{\mathrm{K}}

\newcommand{\s}{\mathrm{s}}

\newcommand{\km}{\;\mathrm{km}}
\newcommand{\yr}{\;\mathrm{yr}}
\newcommand{\Gyr}{\;\mathrm{Gyr}}

\newcommand{\Mpc}{\;\mathrm{Mpc}}

\newcommand{\Msol}{M_{\odot}}

    \def\beq{\begin{equation} }
    \def\eeq{\end{equation} }
    \def\spose#1{\hbox to 0pt{#1\hss}}
    \def\ltsim{\mathrel{\spose{\lower.5ex\hbox{$\mathchar"218$}}
     \raise.4ex\hbox{$\mathchar"13C$}}}

\def\spose#1{\hbox to 0pt{#1\hss}}
\def\lta{\mathrel{\spose{\lower 3pt\hbox{$\mathchar"218$}}
        \raise 2.0pt\hbox{$\mathchar"13C$}}}
\def\gta{\mathrel{\spose{\lower 3pt\hbox{$\mathchar"218$}}
        \raise 2.0pt\hbox{$\mathchar"13E$}}}

\shorttitle{The Assembly of High-z SMBHs}
\shortauthors{Tanaka \& Haiman}

\begin{document}

\title{The Assembly of Supermassive Black Holes at High Redshifts}

\author{Takamitsu Tanaka and Zolt\'an Haiman}

\affil{Department of Astronomy, Columbia University, 550 W120th St., New York, NY 10027}

\begin{abstract}
  The supermassive black holes (SMBHs) massive enough ($\gtrsim$
  few$\times 10^{9}\Msol$) to power the bright redshift $z\approx 6$
  quasars observed in the Sloan Digital Sky Survey (SDSS) are thought
  to have assembled by mergers and/or gas accretion from less massive
  ``seed'' BHs.  If the seeds are the $\sim 10^{2}\Msol$ remnant BHs
  from the first generation of stars, they must be in place well before
  redshift $z=6$, and must avoid being ejected from their parent
  proto--galaxies by the large (several$\times 10^{2} \km\;\s^{-1}$)
  kicks they suffer from gravitational--radiation induced recoil
  during mergers with other BHs. We simulate the SMBH mass function at
  redshift $z>6$ using dark matter (DM) halo merger trees, coupled
  with a prescription for the halo occupation fraction, accretion
  histories, and radial recoil trajectories of the growing BHs.  Our
  purpose is (i) to map out plausible scenarios for successful
  assembly of the $z\approx 6$ quasar BHs by exploring a wide region
  of parameter space, and (ii) to predict the rate of low--frequency
  gravitational wave events detectable by the {\it Laser
  Interferometer Space Antenna} ({\it LISA}) for each such scenario.
  Our main findings are as follows: (1) $\sim 100 \Msol$ seed BHs can
  grow into the SDSS quasar BHs without super--Eddington accretion,
  but only if they form in minihalos at $z\gtrsim30$ and subsequently
  accrete $\gtrsim 60\%$ of the time; (2) the scenarios with optimistic assumptions
  required to explain the SDSS quasar BHs overproduce the mass density
  in lower--mass (few$\times10^{5}\Msol \lsim M_{\rm bh}\lsim$
  few$\times10^{7}\Msol$) BHs by a factor of $10^2-10^3$, unless seeds
  stop forming, or accrete at a severely diminished rates or duty
  cycles (e.g. due to feedback), at $z\ltsim 20-30$.  We also present
  several successful assembly models and their {\it LISA} detection
  rates, including a ``maximal'' model that gives the highest rate
  ($\sim 30$ yr$^{-1}$ at $z=6$) without overproducing the total SMBH
  density.
\end{abstract}

\section{Introduction}

The discovery of bright quasars at redshifts $z\gsim 6$ in the Sloan
Digital Sky Survey (SDSS) implies that black holes (BHs) as massive as
several $\times 10^9{\rm M_\odot}$ were already assembled when the age
of the universe was less than $\approx 1$ Gyr (see the recent review
by Fan 2006).  These objects are among the oldest detected discrete
sources of radiation in the Universe.  The likelihood that all of
these quasars are significantly magnified by gravitational lensing,
without producing detectable multiple images (Richards et al. 2004),
is exceedingly small (Keeton et al. 2005), and if their luminosities
are powered by accretion at or below the Eddington rate, the central
objects must be $\sim 10^{9}\Msol$ supermassive black holes (SMBHs).
In particular, the quasar SDSS J1148+5251 (Fan et al.  2003, 2001) is likely
to be powered by a SMBH with a mass of $\approx 10^{9.5}\Msol$
(Willott et al. 2003).

The mechanism by which such massive BHs formed within 1 Gyr after the
Big Bang remains poorly understood.  Generically, these SMBHs are
thought to have assembled by mergers with other BHs and/or by gas
accretion
\footnote{In this paper, ``accretion'' onto BHs should be assumed to
mean gas accretion, unless otherwise noted.  } onto less massive BHs.
If the first (``seed'') BHs are the $\sim 10^{2}\Msol$ remnant BHs of
the first generation of stars (e.g. Heger et al. 2003), they must be
in place well before redshift $z=6$.  If accretion onto BHs is limited
at the Eddington rate with radiative efficiency $\epsilon$, defined as
the fraction of the rest mass energy of matter falling onto the BH
that is released as radiation, then $1-\epsilon$ of the matter is
accreted and the growth of the BH mass $m$ is given by
\begin{equation}
\frac{d \ln m}{dt}=\frac{1-\epsilon}{\epsilon}\frac{4\pi G\mu m_{\rm p}}
{\sigma_{e}c},
\end{equation}
where $G$ is the gravitational constant, $c$ is the speed of light,
$\mu\approx 1.15$ is the mean atomic weight per electron for a
primordial gas, and $\sigma_{e}$ is the Thompson electron cross
section.  The e--folding time scale for mass growth is $t_{\rm
Edd}\approx 4.4\times 10^7 \;\yr$ for $\epsilon=0.1$.  In the concordance
cosmological model (see below) the time elapsed between redshifts
$z=30$ (when the first seeds may form) and $z=6.4$ (the redshift of
the most distant quasar) is $\approx 0.77$ Gyr, allowing for a mass
growth by a factor of $\approx 10^{7.7}$.  Therefore, individual
$\sim100\Msol$ seeds can grow into the SDSS quasar BHs through gas
accretion alone, provided the accretion is uninterrupted at close to the Eddington
rate and $\epsilon\ltsim 0.1$. A higher efficiency and/or a lower time--averaged
accretion rate will require many seed BHs to merge together; the
number of required mergers increases exponentially for lower
time-averaged accretion rates.

The discovery of the bright quasars at $z\gsim 6$ were followed by the
first successful numerical calculations in full general relativity of
the coalescence of a BH binary and the corresponding emission of
gravitational waves (GWs) (Pretorius 2005; Campanelli et al. 2006;
Baker et al. 2006).  These calculations also confirmed a result
previously known from post-Newtonian (Kidder 1995) and
perturbation-theory treatments (Favata et al. 2004, Schnittman \&
Buonanno 2007): the coalesced product receives a large center-of-mass
recoil imparted by the net linear momentum accumulated by the
asymmetric gravitational wave emission (see Schnittman et al. 2008 for
a recent detailed discussion of the physics of the recoil, and for
further references).  Typical velocities for this gravitational recoil
(or ``kick'') are in excess of $\sim 100 \km~\s^{-1}$. This is likely
more than sufficient to eject the BHs residing in the low--mass
protogalaxies in the early Universe, since the escape velocities from
the DM halos of these galaxies are only a few $\km~\s^{-1}$.

Several recent works have studied the role of gravitational kicks as
an impediment to merger-driven modes of SMBH assembly.  Simple
semi--analytic models show that if every merger resulted in a kick
large enough to remove the seed BHs from halos with velocity
dispersions up to $\approx 50 \km~\s^{-1}$, then super-Eddington accretion
would be required to build SMBHs of the required mass in the available
time (Haiman 2004; Shapiro 2005).  
Monte--Carlo merger tree models that exclude kicks entirely (Bromley et al. 2004)
or which include a distribution of kick velocities extending to low
values (e.g. Yoo \& Miralda-Escud\'{e} 2004 [hereafter YM04]; 
Volonteri \& Rees 2006 [hereafter VR06])
give a slightly
more optimistic picture, showing that if seed BHs form in most
minihalos in the early Universe, and especially if ejected seeds are
rapidly replaced by new seeds (YM04), then SMBH assembly is just
possible before $z\approx 6$ without exceeding the Eddington accretion
rate.  These works are encouraging steps toward demonstrating that
there are plausible physical models that lead to the timely assembly
of SMBHs massive enough to power the $z>6$ SDSS quasars.

At present, we have no direct observational constraints on SMBH
assembly at $z>6$, and there is, in principle, a large range of
``physically plausible'' possibilities.  The {\it Laser Interferometer
Space Antenna} ({\it LISA}) is expected to be able to detect mergers
of SMBHs in the mass range $\sim (10^4$--$10^7)\,{\rm M_\odot}/(1+z)$
out to $z\sim 30$, and to extract binary spins and BH masses with high
precision up to $z\sim 10$ (Vecchio 2004; Lang \& Hughes 2006, 2007).  It is
also likely that by the time {\it LISA} is operational, there will be
additional independent constraints on the demography of high-redshift
SMBHs.  It is therefore a useful exercise to calculate the expected
{\it LISA} event rate from high--redshift SMBH mergers (e.g. Wyithe \&
Loeb 2003; Sesana et al. 2004, 2005, 2007) in a range of plausible
models. Note that published estimates (Menou et al. 2001, Heger et
al. 2003, Menou 2003, Haehnelt 2003,Wyithe \& Loeb 2003, Sesana et al. 2004, Islam
et al. 2004, Koushiappas \& Zentner 2006, Micic et al. 2007,
Lippai et al. 2008, Arun et al. 2008) for the {\it LISA}
event rate, even at lower redshifts, vary by orders of magnitude, from
$\sim 1$ to as high as $\sim 10^4$ yr$^{-1}$; there is a large range
even among models that explicitly fit the evolution of the quasar
luminosity function \citep{lfh08}.  A related open question is to what
degree the {\it LISA} data stream can help pinpoint the actual SMBH
assembly scenario.  One aim of this paper is to understand the model
degeneracies that can lead to similar {\it LISA} data streams.
Another is to explore as much as possible the full variety of {\it
LISA} event rates arising from various ``physically plausible''
assembly models.

The physical factors that determine the growth of SMBHs at high
redshift fall broadly into four categories: (1) the properties of the
initial seed BHs, such as their redshift, mass, and abundance; (2) the
time--averaged gas accretion rate of individual seeds; (3) the merger
rates of BHs; and (4) effects governing the fate of gravitationally
kicked BHs.  The first category determines the number and mass of seed
BHs available for assembly, and depends primarily on the behavior of
gas in the host DM halos, and the mass and metallicity of the first
stars.  The second category measures the subsequent growth through
accretion, and depends on the availability of fuel over the Hubble
time, and its ability to shed angular momentum and accrete onto the
BH. The third category is a combination of the halo merger rate, the
rarity of seeds, and the timescale for the formation, orbital decay,
and ultimate coalescence of a SMBH binary.  Finally, the recoil
velocity of the coalesced binary is determined by the mass ratio and
spin vectors of the BHs, and its subsequent orbit -- and whether it is
retained or ejected before the next merger -- will be determined by
the overall depth of the gravitational potential of the DM host halo,
and on the spatial distribution of gas and DM in the central region of
the halo, which determine drag forces on the kicked BH.

Our approach to model the above effects closely follows those in
earlier works (e.g. YM04; Bromley et al. 2004; Sesana et al. 2004;
VR06): we use Monte Carlo merger trees to track the hierarchical
growth of DM halos, and a simple semi-analytical model to follow the
growth and dynamics of BHs.  We expand over earlier works by adding an
explicit calculation of the orbits of kicked BHs, and
self--consistently include their corresponding time--dependent
accretion rate.  Additionally, we extend the merger tree to redshifts
beyond $z>40$, and we examine a large range of different models.  For
each set of model parameters, we apply our ``tree-plus-orbits''
algorithm to the entire halo population at $z\approx 6$ to construct a
full population of SMBHs at this redshift, and calculate physical
quantities of interest: the mass function, the SMBH-to-halo mass
ratio, the fraction of DM halos hosting SMBHs, and the expected
detection rate of SMBH mergers by {\it LISA}.

This algorithm assembles SMBHs through simple prescriptions of the
aforementioned four categories of physical contributions to SMBH
formation.  We model the seed population by assuming that some
fraction $ f_{\rm seed}$ of DM halos reaching a threshold virial
temperature $T_{\rm seed}$ forms a Pop III remnant BH.  In--between
mergers, the BHs are assumed to accrete gas at a rate of $f_{\rm
duty}$ times the Eddington rate $\dot{m}_{\rm Edd}\equiv
(1-\epsilon)/\epsilon\times L_{\rm Edd}/c^{2}$. Here $f_{\rm duty}$
should be interpreted as the mean gas accretion rate (averaged over
time--scales comparable to the Hubble time) in units of the Eddington
rate.  Note that this prescription makes no distinction between
episodic accretion near the Eddington rate during a fraction $f_{\rm
duty}$ of the time (with no accretion in-between), and constant
accretion at all times at a fraction $f_{\rm duty}$ of the Eddington
rate.  We assume that the mergers of BH binaries closely follow the
mergers of their host halos (but allow for a delay in the latter due
to dynamical friction). Finally, we simulate the orbits of recoiling
BHs under different assumptions about the baryon density profile and
binary spin orientation.  We discuss the relative importance of
assembly model parameters on the final SMBH mass function and the {\it
LISA} data stream, and ask whether {\it LISA} will be able to uniquely
determine the underlying assembly model from data.  We also examine
several variants of the above scenario, in which (i) the seed BHs are massive,
$\sim 10^5 \Msol$, and formed from the super-Eddington accretion
of a collapsed gaseous core; (ii) the DM halo is initially devoid of
gas when the seed BHs is formed; (iii) seed BHs stop forming below
some redshift; and (iv) models which maintain the so--called $M_{\rm
bh}-\sigma$ relation between BH mass and (halo) velocity dispersion
(Ferrarese \& Merritt 2000; Gebhardt et al. 2000) at all redshifts.

This paper is organized as follows.  In \S~2, we detail our
methodology by describing our assumptions, the assembly algorithm,
including the prescriptions of the aforementioned physical effects,
and the different assembly scenarios we consider.  We present and
discuss our main results in \S~3.  In \S~4, we summarize our most
important results, and comment on future prospects to understand SMBH
assembly at high redshift.  To keep our notation as simple as
possible, throughout this paper the capital $M$ will be used to denote
halo masses, and $m$ will refer to BH masses.  In this paper, we adopt
a $\Lambda$CDM cosmology, with the parameters inferred by Komatsu et
al. (2008) using the five-year data from the {\it Wilkinson Microwave
Anisotropy Probe} ({\it WMAP5} ): $\Omega_{\rm CDM}=0.233$,
$\Omega_{b}=0.046$, $\Omega_{\Lambda}=0.721$, $h=0.70$, and
$\sigma_{8}=0.82$.  We use $n_{\rm s}=1$ for the scalar index.

\section{Assumptions and Model Description}

Our strategy is as follows: (1) We use Monte Carlo merger trees to
construct the hierarchical merger history of DM halos with masses
$M>10^{8}\Msol$ at redshift $z=6$, i.e. those that can host SMBHs of
mass $m\gsim 10^{5}\Msol$; (2) We insert seed BHs of mass $m_{\rm
seed}$ into the tree in some fraction $ f_{\rm seed}$ of halos that
reach a threshold temperature $T_{\rm seed}$; (3) We follow the
subsequent BH assembly history by allowing the BHs to grow by gas
accretion in--between mergers, and by calculating the orbit and
accretion history of each recoiling BH in its host halo.  We assume
that BHs add their masses linearly upon merging, and ignore mass
losses due to gravitational radiation, as these losses never
accumulate to significant levels, even through repeated mergers (Menou
\& Haiman 2004).  We prescribe spin distributions of the BHs and gas
distributions within their host halos.  We repeat this procedure for
different mass bins of the halo mass function, until we have a
statistically robust sample to represent the global SMBH mass function
at redshift $z=6$ to an accuracy of a factor of two.  We also record
the BH binary mergers whose masses lie within the mass range $\sim
(10^4$--$10^7)\,{\rm M_\odot}/(1+z)$, corresponding roughly to {\it
LISA}'s sensitivity range.

\subsection{The Merger Tree}

We construct DM merger trees based on the algorithm by Volonteri et
al. (2003), which allows only binary mergers.  Similar numerical
algorithms (e.g. Somerville \& Kolatt 1999; Cole et al. 2000) give
somewhat different results, as have been discussed recently by Zhang
et al. (2008).  We will not reproduce the Volonteri et al. (2003)
recipe here in full; instead we present a brief review. We take the extended Press-Schechter (EPS) mass
function (Press \& Schechter 1974, Lacey \& Cole 1993),
\begin{equation}
\frac{dN}{dM}=\frac{1}{\sqrt{2\pi}}\frac{M_{0}}{M}\frac{d\sigma^{2}_{M}}{dM}
\frac{\delta_{c}(z)-\delta_{c}(z_{0})}{\left(\sigma^{2}_{M}-\sigma^{2}_{M_{0}}\right)^{3/2}}
\exp\left\{-\frac{1}{2}\frac{\left[\delta_{c}(z)-\delta_{c}(z_{0})\right]^{2}}
{\sigma^{2}_{M}-\sigma^{2}_{M_{0}}}\right\},
\end{equation}
which gives for a parent halo of mass $M_{0}$ at redshift $z_{0}$ the
number of progenitor halos in the range $M\pm dM/2$ at a higher
redshift $z$.  Here $\sigma_M$ is the amplitude of the linear matter
fluctuations at redshift $z=0$, smoothed by a top hat window function
whose scale is such that the enclosed mass at the mean density is $M$
(computed using the fitting formula for the transfer function provided
in Eisenstein \& Hu 1999), and $\delta_c$ is the redshift--dependent
critical overdensity for collapse.  We take the limit as $\Delta z
\equiv z-z_{0} \rightarrow 0$:

\begin{equation}
\frac{dN}{dM}=\frac{1}{\sqrt{2\pi}}\frac{M_{0}}{M}\frac{d\sigma^{2}_{M}}{dM}
\;\left(\sigma^{2}_{M}-\sigma^{2}_{M_{0}}\right)^{-3/2}
\;\frac{d\delta_{c}}{dz}
\;\Delta z.
\end{equation}

The two advantages of taking this limit are that (i) by linearizing
the expression, the $z-$ and $M$--dependences separate, allowing a
tabulation as functions of the parent \& progenitor halo mass, and
(ii) separating $\Delta z$ allows for a simple algorithm for adaptive
timesteps to make sure that fragmentations produce binaries at most
(no triplets).

\begin{figure}
\centerline{\hbox{
\plotone{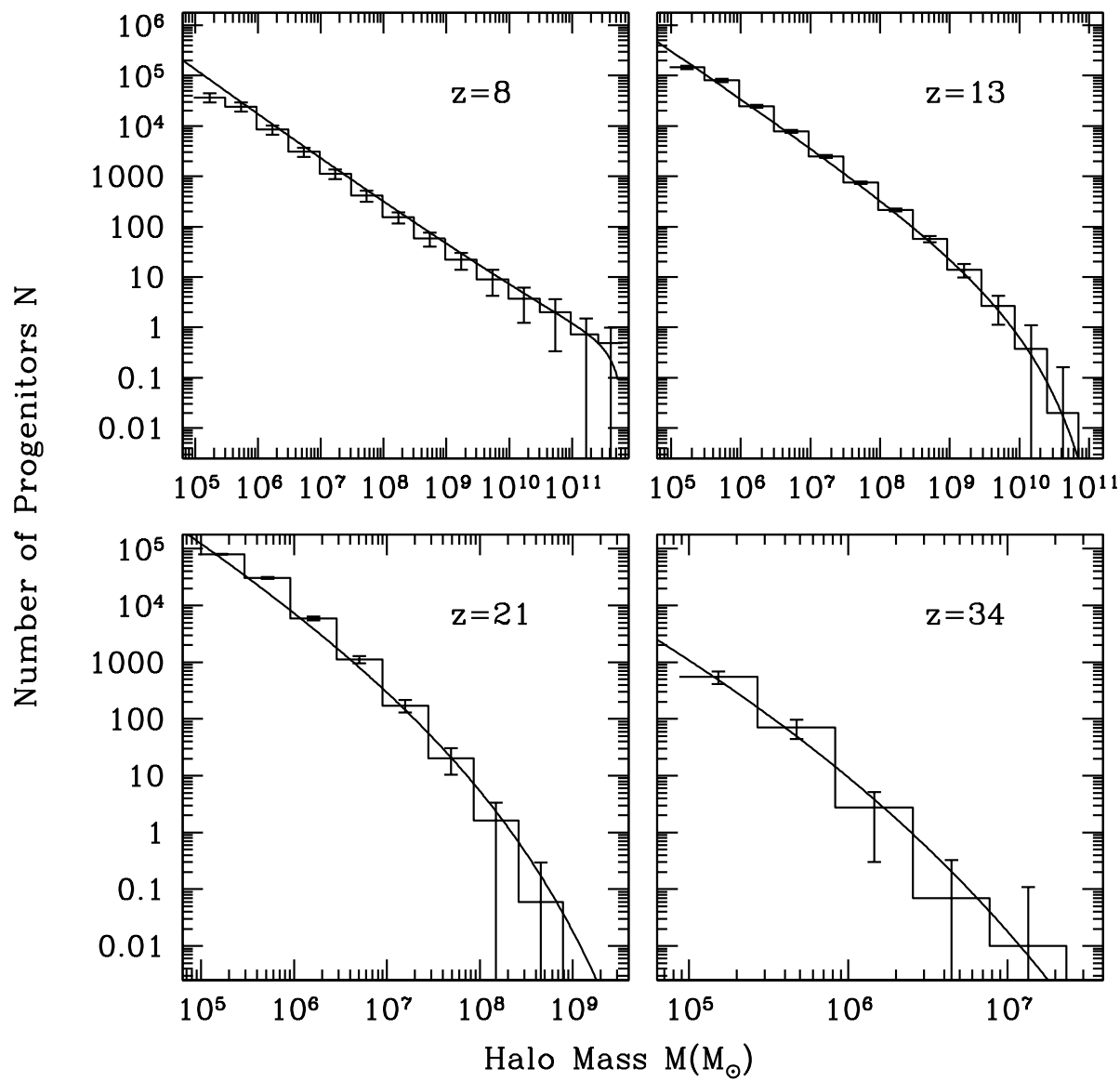}
}}
\caption{The Monte--Carlo--generated mass function of progenitors
for a $10^{12} \Msol$, $z_{0}=6$ parent halo
at $z=8$, 13, 21 and 34.
The histogram is the mean number of 100 realizations,
and the error bars demarcate the Poisson errors.
The solid curve is the EPS prediction.
 }
  \label{fig:tree}
  \end{figure}

For a parent halo of mass $M_{p}$ after a small step $\Delta z$, the
mean number of ``minor'' fragments in the mass range $M_{\rm
lo}<m<M_{p}/2$ is given by
\begin{equation}
\label{PS}
N_{p}=\int_{M_{\rm lo}}^{M_{p}/2} \frac{dN}{dM}dM \propto\Delta z.
\end{equation}
We choose $\Delta z$ adaptively such that $N_{p}\ll1$, which ensures that
multiple fragmentations are unlikely in a given single time step.  We
place a lower limit of $10^{-3}$ (in redshift units) for the timestep
to keep computation times manageable.  The integral in Equation
(\ref{PS}) diverges as $M_{\rm lo}\rightarrow 0$, making it
computationally prohibitive to compute the merger history of
arbitrarily small halos.  To avoid this problem, all progenitors below
a fixed mass resolution $M_{\rm res}$ are considered jointly as
accreted mass and not tracked individually.  Any progenitor with
$M<M_{\rm res}(z)$ is thus discarded from the tree and its prior
history is disregarded.  For our calculations, we choose $M_{\rm
res}(z)$ to be the mass corresponding to a virial temperature of $1200
\K$, corresponding to $M_{\rm res}\sim$ $4.7\times10^{5} \Msol$ at
$z=40$ and $3.4\times10^{6} \Msol$ at $z=10$.  Theoretical studies
have concluded that Pop III stars can start forming at lower virial
temperatures,\footnote{Haiman et al. (1996), Tegmark et al. (1997) and
Machacek et al. (2001) suggest a threshold virial temperature of $\sim
400 \K$ for collapse.  In their recent high-resolution numerical
simulations, O'Shea and Norman (2007) find star formation in halos of
masses $(1.5-7)\times10^{5}\Msol$ between $19<z<33$, with no
significant redshift dependence on the mass scatter.  These values
correspond to virial temperatures of $260-1300 \K$.}  but numerical
considerations have forced us to adopt a somewhat larger threshold.
We do not impose an explicit upper redshift limit, and we run the tree
until our halos at $z=6$ are entirely broken up into progenitors with
$M<M_{\rm res}$.  As a check on our Monte--Carlo merger tree
algorithm, in Figure \ref{fig:tree}, we present the progenitor mass
functions of a $10^{12}\Msol$, $z_{0}=6$ parent halo at redshifts of
$z=8$, $13$, $21$ and $34$, together with the Poisson errors of the 
merger tree output and the
predictions from the EPS conditional mass function (eq.~2).
Our merger tree results are consistent with the EPS conditional mass
function up to redshift $z\approx 40$, with agreement within a factor
of two for most mass bins and redshift values.  In particular, the
numerical mass function agrees well with the EPS prediction for the
low-mass progenitors, even at very high redshit, but the higher-mass
progenitors are under-predicted by a factor of up to two. We note that
Cole et al. (2000) used a numerical algorithm similar to the one we
adopted, to construct a halo merger-tree.  As discussed in Zhang et
al. (2008), that algorithm results in a similar inaccuracy.

\subsection{The Initial Black Hole Population}

The conditions under which the first black holes form are highly
uncertain, though numerical simulations (Abel et al. 2002, O'Shea \&
Norman 2007) do provide useful indications.  We parametrize our
ignorance in terms of a seeding fraction, such that a fraction $f_{\rm
seed}$ of all halos reaching the critical virial temperature $T_{\rm
seed}$ form a seed BH.  There are physical mechanisms that make a low
seeding fraction plausible: the first stars may form only in rare,
baryon-rich overdense regions with unusually low angular momentum, and
seed remnants may receive ejecting kicks from collapse asymmetry
mechanisms similar to those responsible for high-velocity pulsars.
Furthermore, radiative and other feedback processes may prohibit ${\rm
H_2}$--formation, cooling, and star--formation in the majority of
low--mass minihalos at high redshift (e.g. Haiman, Rees \& Loeb 1997;
Mesinger et al. 2006). Since the {\it LISA} event rate, especially at
the earliest epochs, will depend primarily on the abundance of BHs
present, it is highly sensitive to the seeding function.

We choose two fiducial seeding models, the first with $T_{\rm
seed}=1200 {\rm K}$ (the minimum value required for metal-free
molecular line cooling and star formation) for a Pop-III remnant seed
BH with $m_{\rm seed}=100\Msol$.  The second model has $T_{\rm
seed}=1.5\times 10^{4} {\rm K}$ and $m_{\rm seed}=10^{5}\Msol$,
inspired by the ``direct collapse'' models of more massive BHs from
the central gas in halos with a deep enough potential to allow atomic
cooling (Oh \& Haiman 2002; Bromm \& Loeb 2003; Volonteri \& Rees
2005; Begelman et al. 2006; Spaans \& Silk 2006; Lodato \& Natarajan
2006).  If Eddington accretion is the main mode of growth, then we do
not expect the choice of seed mass for each type of model to
qualitatively affect our results, other than the obvious linear
scaling of the overall BH mass function with the initial seed mass.
Only the binary mass ratios affect recoil magnitudes, and the
subsequent orbital dynamics depends minimally on the BH mass.

\subsection{Baryonic and Dark Matter Halo Profiles}

The DM profile for the earliest halos is found to be similar to the
NFW (Navarro, Frenk and White 1997) profile of lower--redshift, more
massive DM halos (Abel et al. 2000; Bromm et al. 2002; Yoshida et
al. 2003).  However, the composition and spatial distribution of the
baryons, at the time when the seed BH appears in these halos, is
poorly understood, and is unconstrained by observations.  This is
unfortunate, since these quantities play a pivotal role in determining
the orbital dynamics and growth rate of a recoiling BH.

A steep profile with a cusp will retain BHs more effectively, owing
both to a deeper gravitational potential well and a larger dynamical
drag force at the halo center.  The baryon distribution will also
determine the accretion history of the central BH by determining the
accretion rate as the BH rests near the halo's potential center, or as
it oscillates in a damped orbit through the halo following a recoil
displacement event.

In addition, whether the baryons are gaseous or stellar has nontrivial
consequences, owing to the difference in the dynamical friction force
between the two cases.  A collisional medium provides a greater
dynamical friction force than a stellar or DM medium with the same
density profile (Ostriker 1999; Escala et al. 2004).  Because of the
difference in the drag force, an environment dominated by gas, and not
by stars (or dark matter), has several possible consequences on BHs:
(1) binaries coalesce more rapidly; (2) a BH recoiling in gas has a
higher likelihood of being retained in its parent halo; and (3) any
``vagrant'' BH that is displaced from the baryon-rich center of the
gravitational potential of its host halo takes less time to return
there, reducing episodes of suppressed accretion.  In
three--dimensional simulations of star--formation in metal--free
minihalos suggest that star--formation is inefficient, with either a
single star, or at most a few stars, forming at the center of the halo
(Abel et al. 2000; Bromm et al. 2002; Yoshida et al. 2003, 2008).
Since in the context of this paper we are concerned with the
pre-reionization Universe, we work with the assumption that stars are
rare before $z\gtrsim6$ and that the baryons in our halos are mostly
gaseous.

We model each galaxy as a spherically symmetric mass distribution with
two components: a DM halo with an NFW profile, and a superimposed
baryonic component.  Previous studies on this subject (see
e.g. Volonteri et al. 2003; Madau \& Quataert 2004) have often assumed
a non-collisional singular isothermal sphere (SIS) profile for the mass
distribution.  This is justified if the gas does not cool significantly
below the virial temperature of the DM halo, and if it has little
angular momentum (so that it is supported thermally, rather than by
rotation).  In most halos whose virial temperature is above $10^4$K,
this assumption is less justified, and a disk may form at the core of
the DM halo (Oh \& Haiman 2002). The direct collapse scenario in
Begelman et al. (2006) and also VR06 adopt such a ``fat disk''
configuration.  However, the central densities of such disks are
within the range of our adopted spherical profiles.  For simplicity,
here we only consider three different prescriptions for spherical gas
distribution.  Our fiducial gas profile is a cuspy, $\rho \propto
r^{-2.2}$ power law, where we have taken the power--law index of $2.2$
as suggested by numerical simulations of the first star-forming
minihalos (Machacek et al. 2001).  This profile is established in
halos that are able to cool their gas via ${\rm H_2}$, and describes
the gas distribution at the time of the first star--formation.

It is possible, however, that the typical seed BHs are surrounded by a
very different gas distribution, at the time of their formation.
First, the progenitor Pop-III stars of the first seed BHs are here
assumed to form in DM halos of mass $\sim 10^{5-6}\Msol$.  The UV
radiation from the star will photo--heat, and easily blow out most of
the gas from low--mass minihalos, even before the star collapses to
leave behind a seed BH (e.g. Whalen et al. 2004).  In this case, the
remnant BH will find itself in a DM halo devoid of gas, and can only
start accreting once a merger with another, gas-rich halo has taken
place, or until the parent halo has accreted enough mass to replenish
its gas (e.g. Alvarez et al. 2008).  We therefore make the simple
assumption that {\it no} gas is present, until the minihalo containing
the newly--formed seed BH merges with another halo, or grows
sufficiently -- assumed here to be a factor of 10 -- in mass.
However, we will examine the consequences of this assumption below, by
performing runs without such a blow--out phase.

Second, as mentioned above, feedback processes may prohibit ${\rm
H_2}$--formation and cooling in the majority of the low--mass
minihalos (e.g. Haiman, Rees \& Loeb 1997; Mesinger et al. 2006). The
gas in such minihalos remains nearly adiabatic, and can not contract
to high densities.  To allow for this possibility, we will study a
variant for the effective gas profile.  Specifically, we adopt the gas
distribution in these halos by the truncated isothermal sphere (TIS)
profile proposed in Shapiro et al. (1999), which has an $r^{-2}$
profile at large radii, but has a flat core at the center owing to the
central gas pressure.  The density profile is normalized (here, and
also in our fiducial gas profile above) such that the total
baryon-to-DM mass ratio inside the virial radius $r_{200}$ equals the
cosmological value.  Both the DM and the baryonic components are
assumed to extend out to $10 r_{\rm vir}$, at which point the density
falls to the background value.  This is consistent with
infall--collapse models of Barkana (2004).

\subsection{Mergers of Dark Matter Halos and Black Holes}

We next have to make important assumptions about the treatment of
mergers between dark matter halos and their resident BHs.

First, we consider the merger between two DM halos, with the more
massive halo referred to as the ``host'' and the less massive as the
``satellite'' halo.  The Press-Schechter formalism and our merger tree
consider as ``merged'' two halos that become closely gravitationally
bound to each other.  However, if the mass ratio of a halo pair is
large, then in reality the smaller halo can end up as a satellite
halo, and its central BH will never merge with that of the more
massive halo.  We therefore require in our models that for BHs in such
halo pairs to coalesce, the halo merger timescale must be shorter than
the Hubble time.  We take the standard parametrization of the merger
time:
\[
\tau_{\rm merge}\approx x\frac{M_{1}}{M_{2}}\tau_{\rm dyn}\approx 0.1 x\frac{M_{1}}{M_{2}}t_{\rm Hub},
\]
where $M_{1}/M_{2}>1$ is the ratio of the halo masses, $x\ltsim 1$ is
some dimensionless factor that encodes the orbit geometry (e.g.
circularity) and dynamical friction, and $\tau_{\rm dyn}$ and $t_{\rm
Hub}$ are the dynamical and Hubble times, respectively.  It has been
suggested (Boylan-Kolchin et al. 2006) that radial infall along
filaments may be preferred in the mergers of elliptical galaxies.
Boylan-Kolchin et al. (2008) give a fitting formula for the merger
time based on numerical simulations.  Their Equation (5) reduces
approximately to $\tau_{\rm merger}/ \tau_{\rm dyn}\approx 0.45$ for a
moderately non-circular orbit with circularity (defined as the ratio of
the orbit's specific angular momentum to the angular momentum of the
circular orbit with the same energy) of $0.5$.  We take a moderate
value of $x\approx0.5$.  That is, if $M_{1}/M_{2}<20$, then the BH in
the smaller halo is considered to be ``stuck'' out in the orbiting
satellite halo and never merges with the central BH of the more
massive halo.  This choice also ensures that the vast majority of BH
binaries in our simulation do not have extreme mass ratios, as the BH
masses co-evolve with the host halos.

We next make assumptions regarding the timescales involving BH
dynamics in their host halos, as follows: (1) if the two merging halos
each contain a central BH, the two holes are assumed to form a binary
efficiently, i.e. we assume there is no delay, in addition to the time
taken by the DM halos to complete their merger; (2) the binary is then
assumed to coalesce in a timely fashion, prior to the interaction with
a third hole; and (3) the binary coalescence is assumed to take place
at the center of the potential of the newly merged halo.  The first
assumption has been addressed by Mayer et al. (2007), who report that
the increased drag force of gas in wet mergers allows the timely
formation of supermassive BH binaries.  The second assumption is valid
for binaries in extremely gas-rich environments (see Milosavljevic \&
Merritt 2003 for a review), or in triaxial galaxies (Berczik et
al. 2006).  As for the third assumption: given that the timescales of
orbital damping are comparable to the intra-merger timescale for BH
velocities of $\gtrsim \sigma_{\rm SIS}$, unperturbed BHs free-falling
during the halo merger process are likely to end up near the center of
the potential within the merging timescales of their hosts.  We do not
include triple-BH interactions in our analysis.

The assumption that BH binaries merge efficiently following the
mergers of their host halos may be unjustified in our models in which
initially, the DM halo is devoid of gas, since gas is generally
believed to be necessary for prompt coalescence.  However, this
inconsistency will not affect our conclusions, for the following
reasons.  First, we find that a BH merger in a gasless environment is
a rare event, as it occurs only if both parent halos are low-mass
halos that had formed seed BHs relatively recently (i.e. neither halos
have yet grown in mass by a factor of 10).  Second, members of such
binaries will have equal (or, in actuality, similar) masses, since
they have not been able to add to their seed mass by accretion.  If
the BHs can merge efficiently without gas, the coalesced product will
likely be ejected, owing to a shallow halo potential and a relatively
large kick of an equal-mass merger.  If the binary does not coalesce
efficiently, it will coalesce once the parent halo merges with a
gas-rich, BH-free halo, or once the parent halo accretes enough gas to
facilitate the merger.  Such belated mergers will also presumably take
place with a mass ratio of close to unity and will likely result in
ejection, regardless of whether significant gas accretion takes place
before coalescence.  Now, consider the case of a stalled binary
encountering a third BH before gas enrichment of the halo.  If the
third BH is much more massive, it will not be ejected by gravitational
interaction or recoil.  There will be a massive BH in the center of
the host halo, and whether the two smaller seed BHs were ejected or
swallowed by the larger BH is of little consequence to our analysis,
especially given the rarity of double-seed binaries.  Thus,
inefficient binary coalescence is of concern only when a double-seed
binary encounters a third BH of comparable mass before the host halo
is gas-enriched.  Such triple-seed systems are likely to be extremely
rare indeed, and unlikely to affect the overall mass function at
$z=6$.  We anticipate that the main effects of a gas-depleted host
halo will be increasing the number of similar-mass mergers following
the initial epoch of seed formation, and slightly reducing the time
available for gas accretion.

\subsection{Gravitational Recoil}

\subsubsection{Probability Distribution of Kick Velocities}

To calculate the recoil velocities of coalesced BHs, we employ the
formulae provided in Baker et al. (2008), which are used to fit their
numerical results,
\begin{eqnarray}
\label{kicks}
v_{\rm recoil}^{2}&=v_{m}^{2}+v_{\perp}^{2}+v_{\parallel}^{2}
+2v_{\perp}\left(v_{m}\cos\xi+v_{\parallel}\sin\xi\right),\\
\nonumber\\
v_{m}&=A\eta^{2}\sqrt{1-4\eta}\left(1+B\eta\right),\\
v_{a^{\parallel}}&=H\frac{\eta^{2}}{1+q}\left(a_{1}\cos\theta_{1}-qa_{2}\cos\theta_{2}\right),\\
v_{a^{\perp}}&=-K\frac{\eta^{3}}{1+q}
\left[a_{1}\sin\theta_{1}\cos\left(\phi_{1}-\Phi_{1}\right)
-qa_{2}\sin\theta_{2}\cos\left(\phi_{2}-\Phi_{2}\right)\right],
\end{eqnarray}
where $q\equiv m_{2}/m_{1}\le1$ is the binary mass ratio, $\eta\equiv
q/(1+q)^{2}$ is the symmetric mass ratio, and $\theta_{1,2}$ are the
angles between the BH spin vectors
$\vec{a}_{1,2}=\vec{S}_{1,2}/m_{1,2}$ and the binary orbital angular momentum
vector.  The angles $\phi_{1,2}$ denote the projection of the spin
vectors onto the orbital plane, measured with respect to a fixed
reference angle.  As seen from the equations themselves, $v_{m}$ is
the kick component that depends only on the symmetric mass ratio;
$v_{a^{\parallel}}$ and $v_{a^{\perp}}$ depend on the mass ratio and
the projection of the binary spins parallel and perpendicular,
respectively, to the orbital angular momentum.
\footnote{Note that this notation differs slightly from Baker et
al. (2008) -- we have simplified their notation to be more transparent
in our spherically symmetric geometry.}  $\Phi_{1}(q)=\Phi_{2}(1/q)$
are constants for a given value of $q$ that encode the dependence of
the kick and orbital precession on the initial spin configuration.  We
use the mean values given in Baker et al. (2008) for the fitting
parameters: $A=1.35\times 10^{4}\km\; \s^{-1}$, $B=-1.48$, $H=7540
\km\; \s^{-1}$, $K=2.4\times 10^{5}\km\; \s^{-1}$, and
$\xi=215^{\circ}$.  We assume spherical symmetry in our host DM halos,
so we are concerned only with the recoil magnitudes and not with the
kick orientations.

Following Schnittman \& Buonanno (2007), for every recoil event, we
assign to both members of the binary spin magnitudes in the range $0.0
\le a_{1,2} \le 0.9$, randomly selected from a uniform distribution.
We consider two scenarios for the spin orientation: a case where the
orientation is completely random, with $0\le \theta_{1,2} \le \pi$ and
$0 \le \phi_{1,2} \le 2\pi$ chosen randomly from a uniform
distribution; and one where the spins are completely aligned with the
orbital angular momentum vector.\footnote{Both cases have the
computational advantage that one does not require the values for
$\Phi_{1,2} (q)$.  For a totally random spin orientation and a given
value of $q$, choosing $\phi_{1,2}$ randomly is equivalent to choosing
$\phi_{1,2}-\Phi_{1,2}$ randomly. When the spins are aligned with the
orbital angular momentum, $\Phi_{1,2}$ terms are irrelevant because
they are always multiplied by $\sin \theta_{1,2}=0$.}  The latter
scenario is motivated by Bogdanovic et al. (2007), who argued that
external torques (such as those provided by a circumbinary accretion
flow) may help align the binary spins with the orbital angular
momentum prior to coalescence, making kicks of $\gtrsim 200
\km\;\s^{-1}$ physically unfavored.  While the argument was originally
used to explain the lack of quasars recoiling along the line of sight
at lower redshifts (Bonning et al. 2007; although see Komossa et
al. 2008 for a candidate recoil detection), the same spin--orbit
alignment will also impact the early assembly history of SMBHs, by
allowing less massive halos to retain recoiling BHs.

Berti \& Volonteri (2008) have provided a merger--tree model to follow
the spin evolution of BHs through gas accretion and binary merger
events.  In this work, we opt not to track the spins of individual BHs
due to the uncertainties involved.  For instance, if circumbinary
disks can act to align the spins of each binary member, this would
present a scenario significantly different from the scenario presented
by Berti \& Volonteri.  As we will show in later sections, the spin
prescription does not appear to play a primary role in determining the
mass function of $z=6$ BHs.

\begin{figure}
\centerline{\hbox{
\plotone{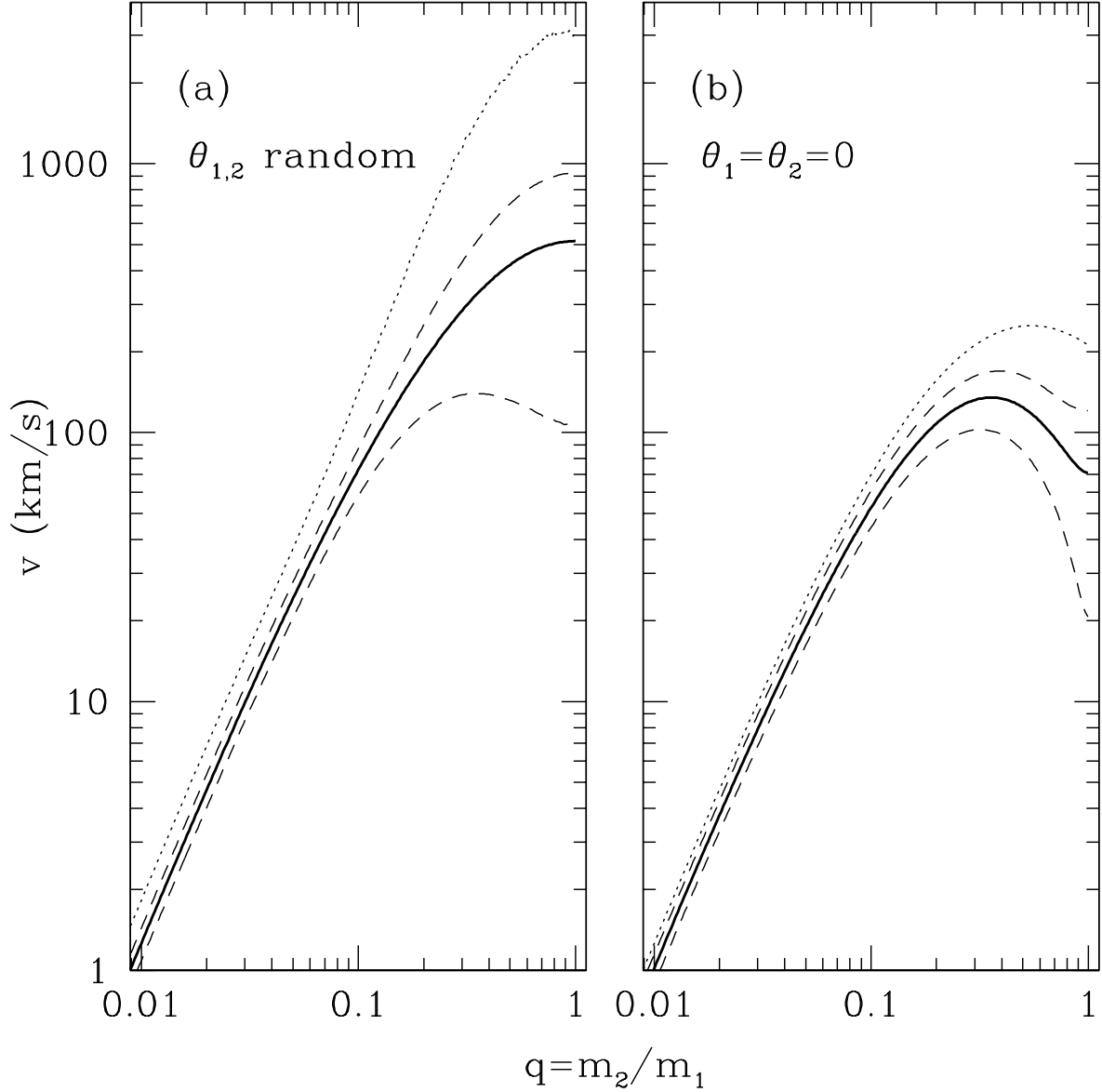}
}}
\caption{The kick velocity distribution as a function of the BH binary
  mass ratio, after $10^{6}$ realizations of Equations (\ref{kicks}-8) at each
  value of $q$.  The left panel shows kicks for random spin
  orientations, while the right panel shows kicks when both BH spins
  are aligned with the orbital angular momentum.  In each case, the
  spin magnitudes are chosen from a uniform random distribution in the
  range $0.0\le a_{1,2} \le 0.9$. The solid curves show the mean, the
  dashed curves show the $1$-$\sigma$ range, and the dotted curve
  gives the maximum value generated in the $10^6$ realizations.  }
  \label{fig:kicks}
  \end{figure}

We show the recoil velocity distribution for both orientation
scenarios as a function of the mass ratio $q$ in Figure
\ref{fig:kicks}.  The figure shows the mean, $1$-$\sigma$, and maximum
values for the recoil velocity magnitude from $10^{6}$ random
realizations at a given value of $0.01\le q\le 1$.  For $q>0.1$, if
spins are randomly oriented then kicks for similar-mass mergers are in
the $100-1000 \km\;\s^{-1}$ range, with a handful of kicks above $1000
\km\;\s^{-1}$ and a maximum possible kick of $\approx 3000
\km\;\s^{-1}$; for spins aligned with the orbital angular momentum,
kicks are typically below $200 \km\; \s^{-1}$, with the maximum
allowed kick no more than $300 \km\;\s^{-1}$.  For $q\gtrsim 0.1$ and
for random spin orientations, the maximum kick $v_{\rm max}(q)$ is
achieved close to where $v_{\parallel}\gg v_{\perp}$ is maximized;
this occurs when $a_{1,2}=0.9$, $\cos(\phi_{1,2}-\Phi_{1,2})=1$, and
$\sin\theta_{1}\approx -\sin\theta_{2}\approx1$.  At these spin
parameter values, $v_{\rm max}(q)$ is well approximated by
$\sqrt{v_{m}^{2}+v_{\parallel}^{2}}$, and is a monotonically
increasing function of $q$.  If the spins are aligned with the orbital
angular momentum vector, then $v_\parallel=0$ and the maximum kick
occurs when $a_{1}=0.0$, $a_{2}=0.9$.  Also, in this aligned case, the
spin-independent component $v_{m}$ is the dominant term for $q\ltsim
0.6$ and peaks at $q\approx 0.345$, resulting in the maximum and mean
values for the recoil speed peaking between these $q$ values.

\subsubsection{Trajectories of Kicked Black Holes}

Given the mass distribution of the host halo and a recoil speed
generated from the method detailed in the previous subsection, we
numerically integrate the equation for the radial motion of a BH with
mass $m$,
\begin{equation}
\label{eq-osc}
\frac{dv}{dt} = 
-\frac{GM(r)}{r^{2}} + a_{\rm DF} - v\frac{\dot m}{m},
\end{equation}
where $r(t)$ is the radial displacement of the BH from the center of
the host halo and $v(t)$ is the BH's radial velocity. The first term
on the right--hand side is the acceleration due to Newtonian gravity
with $M(r)$ the total (dark matter + baryon) mass enclosed inside
spherical radius $r$; the second is the drag deceleration due to
dynamical friction; and the third is the deceleration due to mass
accretion.  A similar calculation of the kicked BH's trajectory has
been performed by Madau \& Quataert (2004) -- the main difference from
our prescription is that they assumed the halo to have a collisionless
SIS profile, and adopted parameters describing galactic bulges in the
local Universe, whereas we use the hybrid DM+gas profile described
above, and adopt parameters relevant to low--mass halos at high
redshifts.

For a non-collisional medium (in our case, for dark matter), the
dynamical friction is described by the standard Chandrasekhar formula
(see e.g. Binney \& Tremaine 1988),
\begin{equation}
a_{\rm DF}^{\rm star}(r, v)=-4\pi G^{2}m\rho(r)\;\frac{1}{v}
\;\ln\Lambda\left[{\rm erf}(X)-\frac{2}{\sqrt{\pi}}X\exp(-X^{2})\right],
\end{equation}
where $\ln \Lambda$ is the Coulomb logarithm and
$X=v/(\sqrt{2}\sigma_{\rm DM})$, with $\sigma_{\rm DM}$ the velocity
dispersion of the DM halo.  In a collisional medium, the density wave
in the wake of an object traveling at near or above the sound speed is
enhanced via resonance, an effect that has no counterpart in
collisionless media.  This results in an enhancement of the dynamical
friction force, for which Ostriker (1999) has derived an analytical
formula.  However, the Ostriker prescription is known to overpredict
the drag force at slightly supersonic velocities when compared with
numerical simulations.  While Escala et al. (2004) provides a fitting
formula that better fits the numerical results at low speeds, their
formula suffers from the opposite problem, and over-predicts the drag
for highly supersonic motion.  We therefore adopt a hybrid
prescription, adopting the Escala et al. (2004) formula for motion with
$\mathcal{M}<\mathcal{M}_{eq}$ and the Ostriker formula for
$\mathcal{M}>\mathcal{M}_{eq}$, where $\mathcal{M}_{eq}$ is the Mach
number ($=|v|/c_{s}$) where the two prescriptions predict equal drag.
The entire prescription is described by
\begin{equation}
a_{\rm DF}^{\rm gas}(r, v)=-4\pi G^{2}m\rho(r)\frac{1}{v}
\;\times\; f(\mathcal{M})
,\qquad{\rm where:}
\end{equation}
\begin{equation}
f(\mathcal{M})=
\left\{ \begin{array}{ll}
       0.5\ln\Lambda\left[{\rm erf}\left(\frac{\mathcal{M}}{\sqrt{2}} \right)-\sqrt{\frac{2}{\pi}}
       \mathcal{M}\exp(-\mathcal{M}^{2}/2)\right]
         & \mbox{if $ \;\;\;0\leq\mathcal{M} \leq 0.8$};\\
       1.5\ln\Lambda\left[{\rm erf}\left(\frac{\mathcal{M}}{\sqrt{2}} \right)-\sqrt{\frac{2}{\pi}}
       \mathcal{M}\exp(-\mathcal{M}^{2}/2)\right]
          & \mbox{if $ 0.8<\mathcal{M} \leq \mathcal{M}_{eq}$};\\
        \frac{1}{2}\ln\left(\frac{\mathcal{M} +1}{\mathcal{M} -1}\right)-\ln\Lambda
        & \mbox{if $\;\;\;\;\;\;\;\;\mathcal{M}>\mathcal{M}_{eq}$}.\end{array} \right. 
\end{equation}
The Coulomb logarithm $\ln \Lambda$ is not a precisely known
parameter, but is generally agreed to be $\gtrsim 1$ for both the
stellar and the gaseous cases.  We adopt the value $\ln \Lambda=3.1$
used in Escala et al. (2004), which yields
$\mathcal{M}_{eq}\approx1.5$.

The drag force depends on the local gas sound speed.  Instead of
attempting to compute a temperature profile explicitly, we make the
approximation that the gas sound speed is constant and given by the
isothermal sound speed of the halo virial temperature.  We believe
this to be justified from numerical simulation results that show the
gas temperature to vary by at most a factor of $\approx 3$ within the
virial radius despite a steeper-than-isothermal ($\propto r^{-2.2}$)
density profile (see, e.g. Machacek et al. 2001).  While local
variations in the sound speed may have significant effects when $v\sim
c_{s}$, the recoil events of interest here are for the most part
highly supersonic.  Recoil events with $v\sim c_{s}$ will result in
quick damping of the BH orbit and for the purposes of this paper will
in all likelihood be indistinguishable from the zero-recoil
calculation in terms of their accretion history.

The virial temperature is given by (e.g. Barkana \& Loeb 2001)
\begin{equation}
T_{\rm vir}\approx 370(1+z)\left(\frac{\mu}{0.6}\right)
\left(\frac{M}{10^{7}\Msol}\right)^{2/3}\left(\frac{\Omega_{0}h^{2}}{0.14}\right)^{1/3} \K ,
\end{equation}
and the isothermal sound speed is
\begin{equation}
c_{s}\approx1.8\;(1+z)^{1/2}\left(\frac{M}{10^{7}\Msol}\right)^{1/3} \left(\frac{\Omega_{0}h^{2}}{0.14}\right)^{1/6}\km\;{\rm s}^{-1}
\end{equation}
We also employ a simplified prescription for the velocity dispersion
of non-collisional matter, given by the SIS expression evaluated at the
virial radius: $\sigma_{\rm SIS}=\sqrt{GM/2r_{200}}$.  The simplified
expression agrees with the exact velocity dispersion for the NFW
profile within $\sim 20\%$ inside the virial radius.

Because matter that is bound to the BH does not contribute to
dynamical friction, we follow Madau \& Quataert (2004) and truncate
the density profiles at the BH radius of influence, $r_{\rm BH}\approx
Gm/\sigma_{\rm SIS}^{2}$.  The density is furthermore assumed to be
constant inside this radius.  Although the BH will drag with it the
surrounding gravitationally bound matter, effectively increasing its
mass, the additional mass is small owing to the large initial recoil
velocity (e.g. Lippai, Frei \& Haiman 2008), and we have ignored this
mass--enhancement here.

\begin{figure}
\centerline{\hbox{
\plotone{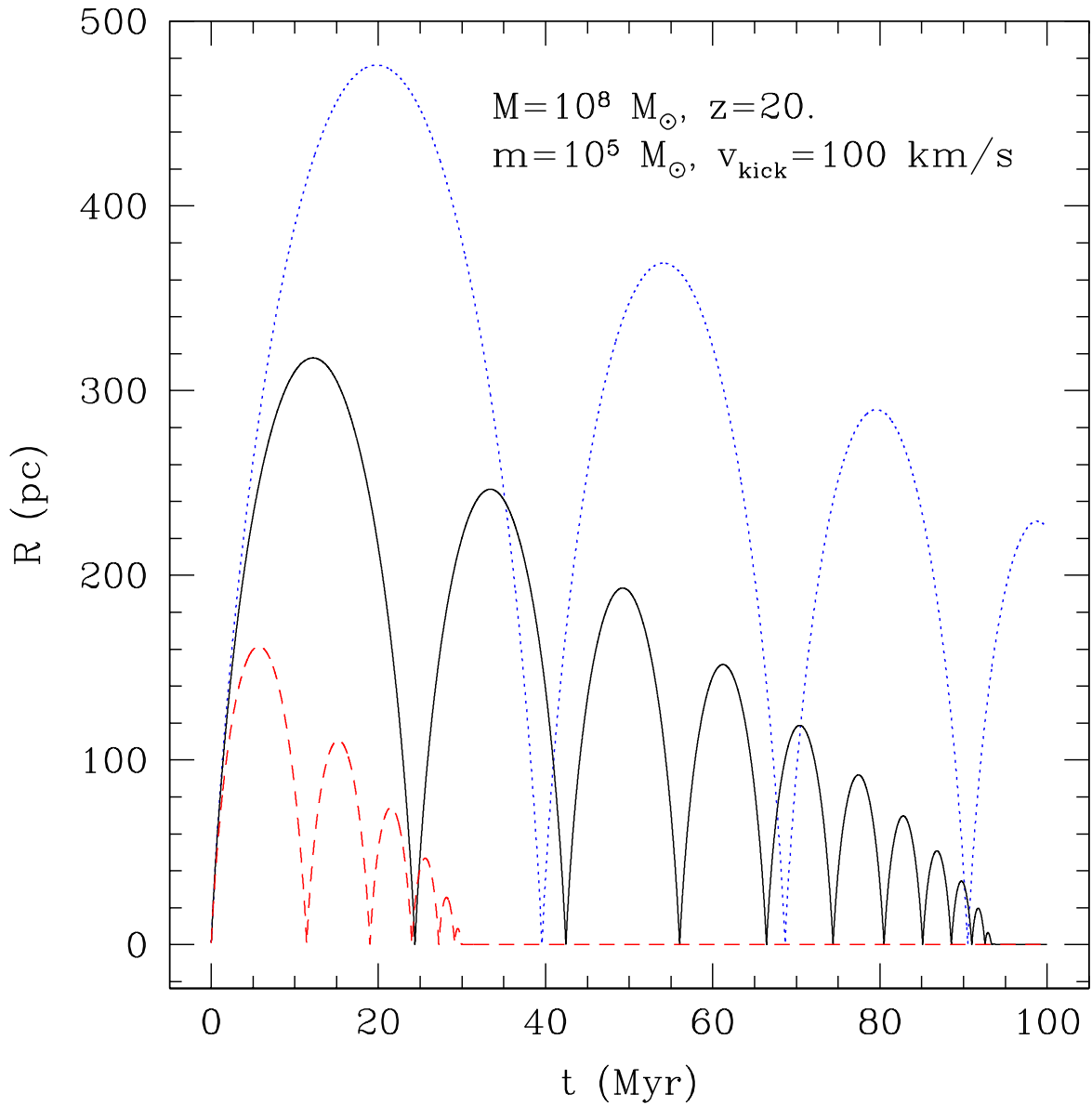}
}}
  \caption{Examples of the radial motion of a recoiling black hole,
for three different mass profiles for the host halo.  The black curve shows the
motion in a pure NFW halo; the halo of the blue (dotted) curve assumes that the
host galaxy has a DM halo with an NFW profile and a corey gas
component; the red (dashed) curve assumes a DM halo and a cuspy $r^{-2.2}$
power law gas profile.  In all cases, the halo mass is $10^{8}\Msol$,
the BH mass is $10^{5}\Msol$, the redshift is $z=20$ and the kick
velocity is $100 \km~\s^{-1}$. }
  \label{fig:osc}
  \end{figure}

Figure \ref{fig:osc} illustrates the effect of the halo matter
distribution on the BH orbit.  For each of the three orbits shown in
the Figure, we adopt $M=10^{8} \Msol$, $m=10^{5}\Msol$, $z=20$ and
$v_{\rm kick}=100 \km~\s^{-1}$.  The black curve shows, for reference,
the radial orbit for a pure NFW profile.  The red curve corresponds to
the case which includes an NFW DM component and a power-law gas
profile with $\rho\propto r^{-2.2}$.  The blue curve is for a halo
with an NFW DM component and a TIS gas profile.

In our calculations, we are mainly interested in whether the kicked BH
is ejected and lost (i.e., can not contribute to the final SMBH mass
at $z=6$), or is retained (i.e., eventually returns to the nucleus,
and can be incorporated into the $z=6$ SMBH).  We place the following
retention condition for recoiling BHs: the BH must return to within
1/10th of the virial radius of the newly merged host halo within
1/10th of the Hubble time.  The fate of a (SM)BH placed in an orbit
extending to the outskirts of its host halo is uncertain: even if it
is not lost through tidal interactions with an incoming merging halo,
it is not likely to form a binary that hardens efficiently.  We
therefore impose the above conservative cutoff, in order to avoid
tracking these vagrant BHs.  Our retention threshold velocity, $v_{\rm
ret}$, above which recoiling BHs do not return within the prescribed
time limit and are considered lost, is a numerically calculable
function of $M$, $m$, $z$ and halo composition.  In order to minimize
computation time, we tabulate $v_{\rm ret}$ in the range $5<z<40$,
$10^{5}\Msol<M<10^{15}\Msol$ and $10^{-6}<m/M<1$ and approximate the
result with a fitting formula that accurately reproduces the exact
numerical results within $5\%$ in the tabulated range.  In principle,
we should compute the retention velocity as a function of the time to
the next merger experienced by the halo.  However, we find the time
dependence to be weak.  The distribution of the time intervals between
halo mergers in a given merger tree has a sharp peak at $\approx
10^{-1} t_{\rm Hub}$, with far fewer mergers occurring at $\sim 10^{-2}
t_{\rm Hub}$ and $\sim t_{\rm Hub}$.  At these tails of the
distribution, $v_{\rm ret}$ varies by $\ltsim 10\%$ from the value for
$10^{-1} t_{\rm Hub}$.  We find that $v_{\rm ret}\sim 5-8 \times
\sigma_{\rm SIS}$.  This is comparable to the escape velocity for a
{\it non-dissipative} pure SIS profile that is truncated at the BH
radius of influence, $v_{\rm esc} \approx 5\sigma_{\rm SIS}$ if
$m=10^{-3}M$ (YM04).

The weak dependence on $v_{\rm ret}$ on the return time limit is a
counterintuitive result, but it can be understood as follows.  There
is a minimum kick speed that is required to displace the hole beyond
$0.1 r_{200}$, which represents $\lim_{t\to0}v_{\rm ret}$; and there
is a maximum kick, $\lim_{t\to\infty}v_{\rm ret}$, beyond which the BH
remains completely unbound from the halo, even in the presence of
drag.  $v_{\rm ret}(t)$, then is a function of time that is always
between these two extreme values.  However, owing to the high central
density of our gas--dominated halos, the difference between these two
limits is small, $\sim 10\%$.  Since this difference is smaller than
other model uncertainties, such as those stemming from discrepancies
from the actual density profile of high-redshift DM halos (e.g.
clumping or triaxiality) and the merger tree prescription, we simply
use the retention velocity computed for the approximate median time
limit, $0.1t_{\rm Hub}$.

Finally, for simplicity, we treat the halo as a static mass
distribution during each recoil event.  That is, we ignore the
cosmological evolution of the DM potential, and we assume that the
recoiling BH does not affect the medium by clumping or heating it.
Note, however that the latter feedback may play a nontrivial role in
real systems, since the kinetic energy of a recoiling hole can be
comparable to the gravitational binding energy of the entire host halo
and can be expected to cause significant disruption of the surrounding
matter.

\subsection{The Black Hole Accretion Rate}

We turn now to our prescription for the accretion rate of the BHs in
our model.  Of particular interest is the possibility that the
gravitational recoil effect will significantly limit the ability of
kicked BHs to accrete gas, by displacing them into low-density regions
for prolonged periods, and/or by limiting through high relative
velocities the amount of gas that can be gravitationally captured.
One can imagine a scenario in which a SMBH whose progenitors have
survived numerous kicks but have spent long episodes in underdense
regions may have a final mass much less than that predicted by simple
Eddington growth.  We therefore follow the accretion rate
self-consistently, as the recoiling holes proceed along their radial
orbits.  Specifically, in our models a BH embedded in gas is assumed
to undergo standard Bondi-Hoyle-Littleton (BHL) accretion,
 \begin{equation}
 \label{Bondi}
 \dot m(r,v) = \frac{4\pi G^{2}\rho_{b}(r)} {(c_{s}^{2}+v^{2})^{3/2}}m^{2}.
\end{equation}
The accretion rate is capped at the Eddington rate,
\begin{equation}
\label{Edd}
\dot m=\frac{1-\epsilon}{\epsilon}\frac{m}{t_{\rm Edd}}, 
\end{equation} 
where $t_{\rm Edd}=44 {\rm Myr}$ and $\epsilon$ is the radiative
efficiency, for which we assume $\epsilon = 0.10$.

If the gas density is too low, or if the sound speed or its velocity
with respect to the gas disk too high, a BH may not be able to accrete
at the Eddington rate even if it is close to the center of a halo.
Because the BHL rate is proportional to $m^{2}$, an underfed BH with
initial mass $m_{0}$ will eventually reach the Eddington accretion
rate ($\propto m$) at a threshold mass
\begin{equation}
m_{\rm Edd}=\frac{1-\epsilon}{\epsilon}\frac{c_{s}^{3}}{4\pi G^{2}\rho_{b\;} t_{\rm Edd}}
\approx3500\left(\frac{c_{s}}{4 \km\;{\rm s}^{-1}}\right)^{3}
\left(\frac{\rho_{b}}{\Msol \rm{pc}^{-3}}\right)^{-1} \Msol,
\end{equation}
where $4 \km\;{\rm s}^{-1}$ is the isothermal sound speed for an
ionized hydrogen gas at $1200\K$.  The typical central density of a
$1200 \K$ halo with a TIS gas profile is $\sim 5\times 10^{-3}
(1+z)^{7/3} \Msol \rm{pc}^{-3}$.  Sub-Eddington accretion rates are
not an issue for BHs in a power-law gas profile, as the steep profile
provides a sufficient central density for immediate Eddington
accretion.

The time it takes for a BH with $m_{0}<m_{\rm Edd}$ to reach this
threshold mass, assuming that the local density remains constant, is
\begin{equation}
t_{\rm crit} =
\left(\frac{m_{\rm Edd}}{m}-1\right) \frac{\epsilon}{1-\epsilon} t_{\rm Edd}
\approx 2.7\left(\frac{m_{0}}{100\Msol}\right)^{-1}
\left(\frac{c_{s}}{4 \km\;{\rm s}^{-1}}\right)^{3}
\left(\frac{\rho_{b}}{\Msol \rm{pc}^{-3}}\right)^{-1} {\rm Gyr}.
\end{equation}
$t_{\rm crit}\sim {\rm a\; few\;} 100 \;{\rm Myr}$ for $100 \Msol$
halos embedded in $1200 \K$ TIS halos at $z\gtrsim 20$, and $t_{\rm
crit}\gtrsim \Gyr$ for $z\ltsim 14$.  This means that for TIS gas
profiles, seed holes will spend a significant fraction or all of the
available time prior to $z\approx 6$ accreting below the Eddington
rate.

The difference between BHL and Eddington accretion rates as they
relate to BH growth is also discussed in Volonteri \& Rees (2006).
However, in that paper the context is for the direct formation of
$m>10^{4}\Msol$ intermediate-mass BHs through super-Eddington BHL
accretion.  We here adopt the opposite extreme assumption, i.e. that
the BH radiates efficiently at all times, and its accretion rate obeys
the Eddington limit. The BHL rate can then initially be sub-Eddington
in TIS halos, owing to the low BH mass and low gas density (the baryon
density required to fuel BHL accretion at the Eddington rate was also
discussed by Turner 1991).

\begin{figure}
\centerline{\hbox{
\plotone{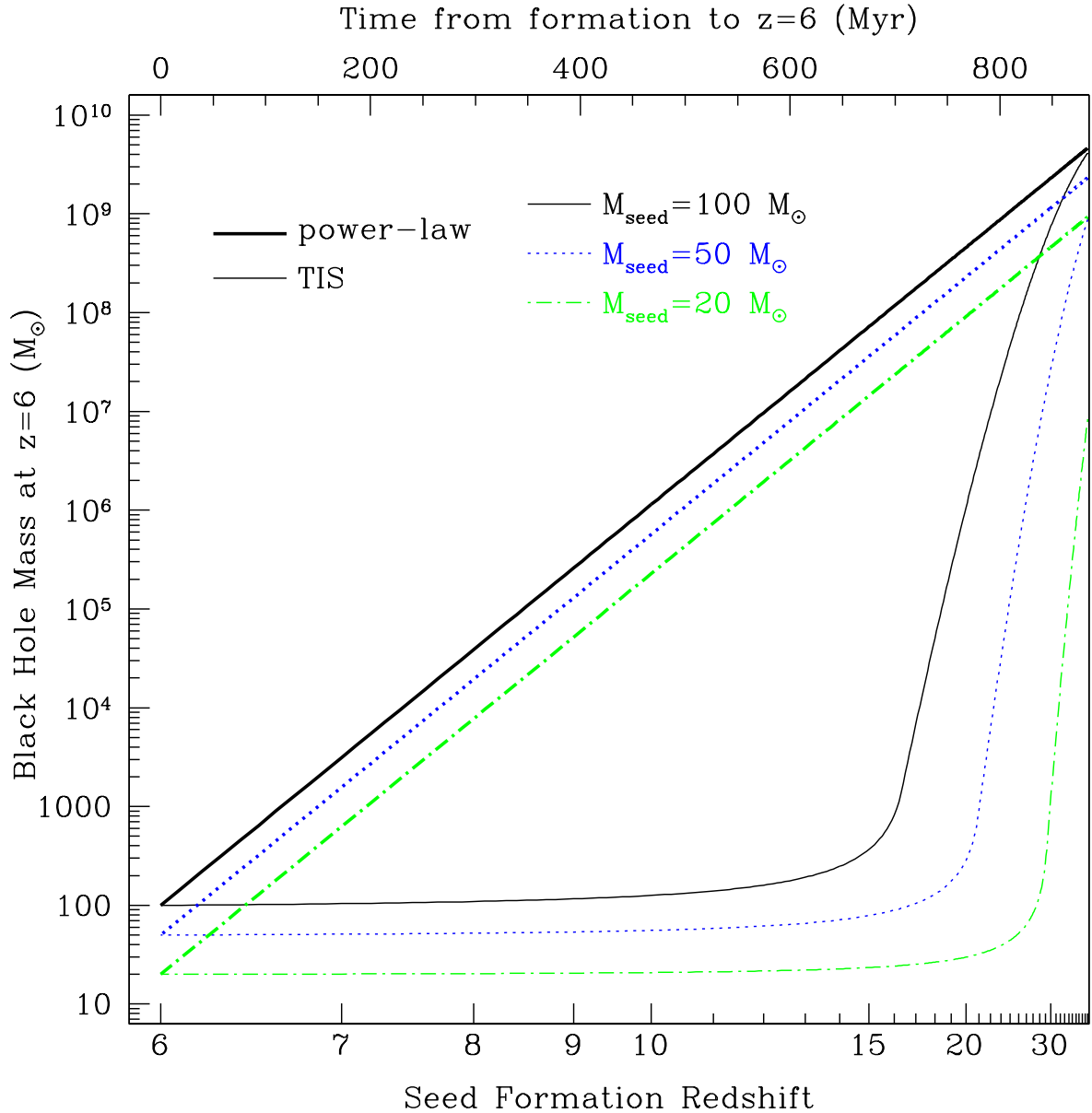}
}}
  \caption{The maximum possible accreted mass by redshift $z=6$ for a
seed BH born with a seed mass $m_{\rm seed}$ in a halo with virial
temperature $T_{\rm vir}=1200 \K$ at redshift $z$.  
If the BH is always surrounded by a steep gas profile, with a power-law cusp, the growth remains Eddington throughout
(which appears as a straight line in this log-log plot). If the gas profile has a flat core (as in the
TIS profile), the central density is initially insufficient to feed
the BH at the Eddington rate, resulting in much slower growth.  The cuspy and corey gas distributions
are demarcated by thick and thin lines, respectively, and different seed BH masses are shown
in different colors (line styles).}
  \label{fig:acc}
  \end{figure}

To illustrate the impact of extended sub--Eddington growth phases, we
perform a simple analytical calculation.  Suppose that a BH is formed
with mass $m_{\rm seed}$ at redshift $z$ in a halo with virial
temperature $1200 \K$, and that the local gas density is held constant
at the value when the BH was formed.  Note that for this exercise, we
assume that gas density is constant even as the halo around the BH is
growing by merging with other halos -- in other words, we assume these
mergers deliver gas to the nucleus containing the original BH, roughly
maintaining a constant density at the Bondi radius around the BH.
Figure \ref{fig:acc} shows the maximum possible mass that can be
attained by such a BH growing in isolation through gas accretion
alone, assuming that the accretion rate is determined solely by
Equations (\ref{Bondi}) and (\ref{Edd}) and that accretion is not
supply--limited.  If the host halo has a steep power-law cusp, the
accretion rate is Eddington throughout regardless of when the seed BH
is formed.  However, if the central fuel density is low, then it is
possible for the local BHL accretion rate to be significantly
sub-Eddington initially.  In such a scenario, the earliest forming
seeds are the only ones able to reach the Eddington rate; the
late-forming seeds are unable to reach the Eddington rate before
$z=6$.  In this scenario the late-forming seeds, which are easily
identifiable by the drastically shallower growth slope in the figure,
cannot grow rapidly enough to contribute to the SMBH population.  Note
that assuming a constant gas density in this calculation gives an
optimistic accretion rate for the TIS case, as in those profiles the
central gas density generally decreases with Hubble expansion and
significantly reduce the BHL rate.

It is computationally expensive to numerically integrate the
individual orbits and accretion histories of every recoiling BH in our
simulations.  We therefore tabulate the accretion growth in Eddington
units during the first $0.1t_{\rm Hub}$ of the orbit, and describe the
results in a fitting formula in the same manner as we have done for
the retention velocity.  Given a specific prescription for the baryon
distribution, we tabulate across four relevant variables in the
following ranges: $10^{5}\Msol<M<10^{15}\Msol$, $10^{-6}<m/M<1$,
$5<z<40$, and $0<v_{\rm kick}/v_{\rm ret}<1$.
  
In the absence of a kick, and if the accretion rate were always at the
Eddington limit, the SMBH mass in a given halo at $z\approx6$ is
easily approximated by
\begin{equation}
\label{est}
m_{\rm SMBH}\approx N_{\rm seed}\;
\exp\left(f_{\rm duty}\frac{\epsilon}{1-\epsilon}\frac{t_{{\rm
seed},6}}{t_{\rm Edd}}\right)m_{\rm seed},
\end{equation}
where $m_{\rm seed}$ is the seed BH mass, $N_{\rm seed}$ is the number
of seeds in the merger history of the halo, $t_{{\rm seed},6}$ is the
available time between the typical seed formation time and $z\approx6$
and $f_{\rm duty}$ is the time--averaged duty cycle for accretion.
Equation (\ref{est}) represents the ideal, maximally efficient
scenario for SMBH assembly, and we can use it to effectively measure
the cumulative impact of underfed accretion, recoil-induced ejection,
and other factors that limit the assembly efficiency.

The measurements of clustering of quasars in the SDSS suggest that the
duty cycle of active (luminous) accretion increases steeply with
redshift at $3\ltsim z\ltsim 6$, with the most active quasar BHs at
$z\approx 6$ showing $0.6 \lsim f_{\rm duty}\lsim 0.9$ (Shen et
al. 2007; Shankar et al. 2008).  We therefore adopt duty cycles of
$\ge 0.6$.  Although it is likely that SMBHs regulate their own growth
through feedback mechanisms, we do not address such scenarios a
priori.  To keep our models as simple as possible, we will only impose
the loosest upper limit on the SMBH accretion rate: they must not
accrete more mass than there the total mass of baryons in the host
halo.  However, we will discuss an alternative ad--hoc model below,
which is able to reproduce the well-known relation between SMBHs and
the velocity dispersions of the bulges of host galaxies (the
$m$-$\sigma$ relation).

\subsection{Putting Together the $z=6$ SMBH Mass Function}

Explicitly constructing the SMBH mass function at $z=6$ over a wide
mass range is computationally intractable.  The host halo mass
inferred from the observed quasar space density from the $z\sim 6$
quasars is several $\times 10^{12}\Msol$ (e.g. Fan 2006).  For every
halo with this mass, there are $\sim 10^{7}$ halos with $10^{8}-
10^{9}\Msol$.  Blindly calculating the SMBH mass for every halo with
$M>10^{8}\Msol$ using our trees-plus-orbits algorithm would be
prohibitively expensive computationally.

We therefore carry out a piecewise calculation of the SMBH mass
function that is computationally tractable.  The procedure is as
follows: (1) We group the halo population into logarithmic mass bins
of size $x<\log_{10}M<x+\Delta x$; (2) For each bin, we select
$\gtrsim 10^{2}-10^{4}$ individual Monte-Carlo-generated halos with masses randomly generated
from the Press-Schechter distribution at $z=6$; (3) We simulate the BH
population for each such halo using our trees-plus-orbits algorithm,
and assume the resulting sample is representative of all $z=6$ halos
in the mass bin; (4) For each bin we multiply the sample by the
appropriate weight to construct the entire Press-Schechter halo mass
function, $\int_{x}^{x+\Delta x} dN/d\ln M d \ln M$; and finally (5) Sum the
contributions from each bin.  The result is a Press-Schechter
distribution of halos with $M>10^{8}\Msol$ at $z=6$, with a
statistical representation of the corresponding SMBH mass function.
The bins used and their relevant properties, including the number of
Monte-Carlo halos that were cloned to populate the full mass function, are
listed in Table 1.  This numerical shortcut is not used for the most
massive halos.  40 halos are expected above $10^{12.85}\Msol$, and
these are simulated individually.

This method allows a fast calculation of the BH mass function, with
the caveat that there must be enough BHs in each bin to keep
statistical uncertainties to a minimum.  Unfortunately, this is not
always preventable for models with extremely low $ f_{\rm seed}$, and
the reader will notice statistical noise in the results of such
models.  In some cases, we increase the halo sample by a factor of 10
in an attempt to reduce the errors.

In all, our simulations represent the Press-Schechter population of DM
halos in a comoving volume of $\approx 280 {\rm Gpc}^{3}$, roughly
equal to the comoving volume that was probed by the SDSS 
between $z\approx 5.7-6.4$.  We have simulated a total of $\approx 1.17
\times 10^{5}$ DM halos (see the column $N_{\rm sim}$ in Table 1),
and through the procedure described above
extrapolated the results to represent the $\approx 3\times 10^{12}$ Press-Schechter halos
(with $M>10^{8}\Msol$)
expected to be present in the $280 {\rm Gpc}^{3}$ volume.

\section{Results and Discussion}

\subsection{Building the $> 10^9{\rm M_\odot}$ SMBHs}

As stated in the Introduction, our primary goals are (i) to understand
the possible ways in which the $> 10^9{\rm M_\odot}$ SMBHs may have
been assembled at redshift $z>6$, and (ii) whether the {\it LISA}
event rates are sufficiently different in competing models so that one
can disentangle the physical assembly scenario from the {\it LISA}
data stream. As a first step toward these goals, we would like to
survey all feasible combinations of the physical assembly parameters,
understand the impact of each model parameter, and look for
corresponding give--away features in the predicted {\it LISA} stream.

Although a broad simulation survey is beyond the scope of this paper,
we have undertaken a cursory tour of the basic parameter space.  We
begin by considering two basic seed models: $100 \Msol$ seeds forming
as remnants of Pop III stars in minihalos when they first reach the
virial temperature $T_{\rm vir}\geq T_{\rm seed}=1200 \K$ and
$10^{5}\Msol$ seeds forming as a result of direct collapse of gas in
halos when they first reach $T_{\rm vir}\geq 1.5\times 10^{4}\K$.  For
each case, we consider halos with an NFW DM component and a gaseous
component with either a cuspy $\rho_{\rm gas}\propto r^{-2.2}$
power-law or a corey TIS profile.  The TIS models consistently failed
to produce SMBHs by $z=6$, with typical maximum BH masses of $\ltsim
10^{3}\Msol$.  In those models the central gas densities are too low
to allow for prolonged episodes of accretion near the Eddington rate,
as noted in Section 2.5 above.

For each type of seed, we vary three sets of parameters: (i) the
seeding fraction $10^{-3}\leq f_{\rm seed}\leq 1$, i.e. the
probability that a halo reaching the critical temperature will form a
seed BH; (ii) the time--averaged accretion rate in Eddington units,
characterized by the duty cycle $f_{\rm duty}$, for which we use
$f_{\rm duty}=0.6$, $0.8$ and $1.0$ (note that $f_{\rm duty}$ and the
radiative efficiency $\epsilon$ are degenerate in our prescription;
see below); and (iii) whether at the time of their merger, the BH
spins are randomly oriented or aligned with the orbital angular
momentum of the binary.  The spin magnitudes are chosen from a uniform
random distribution $0<a_{1,2}<0.9$ (Schnittman \& Buonanno 2007).  We
do not track the evolution of BH spins in our models.  While more
rapidly spinning BHs are capable of accreting at higher efficiency, we
neglect this effect and assume a global efficiency coefficient in our
models.  Of the four main ingredients of the SMBH assembly introduced
in \S~1 above, we here therefore vary three: the seed rarity, the
accretion rate, and the recoil dynamics.  For the fourth, the merger
rate, we have simply assumed that BHs merge when their parent halos
do, as extreme--mass BH binaries are rare in our simulation, given the
threshold we have imposed on the mass ratio for halo mergers.  We will
call the above our fiducial set of parameters.

For each realization, we compute the mass function of BHs at $z=6$
with $m>10^{5}\Msol$; the rate of events in the {\it LISA}
observational mass range, $\sim 10^{4-7}/(1+z)\Msol$; the fraction of
DM halos hosting a central massive BH; and $m/M$, the ratio between
the mass of the SMBH and its host DM halo, which serves as a proxy for
the $m$-$\sigma$ relation.

\begin{figure}
\centerline{\hbox{
\plotone{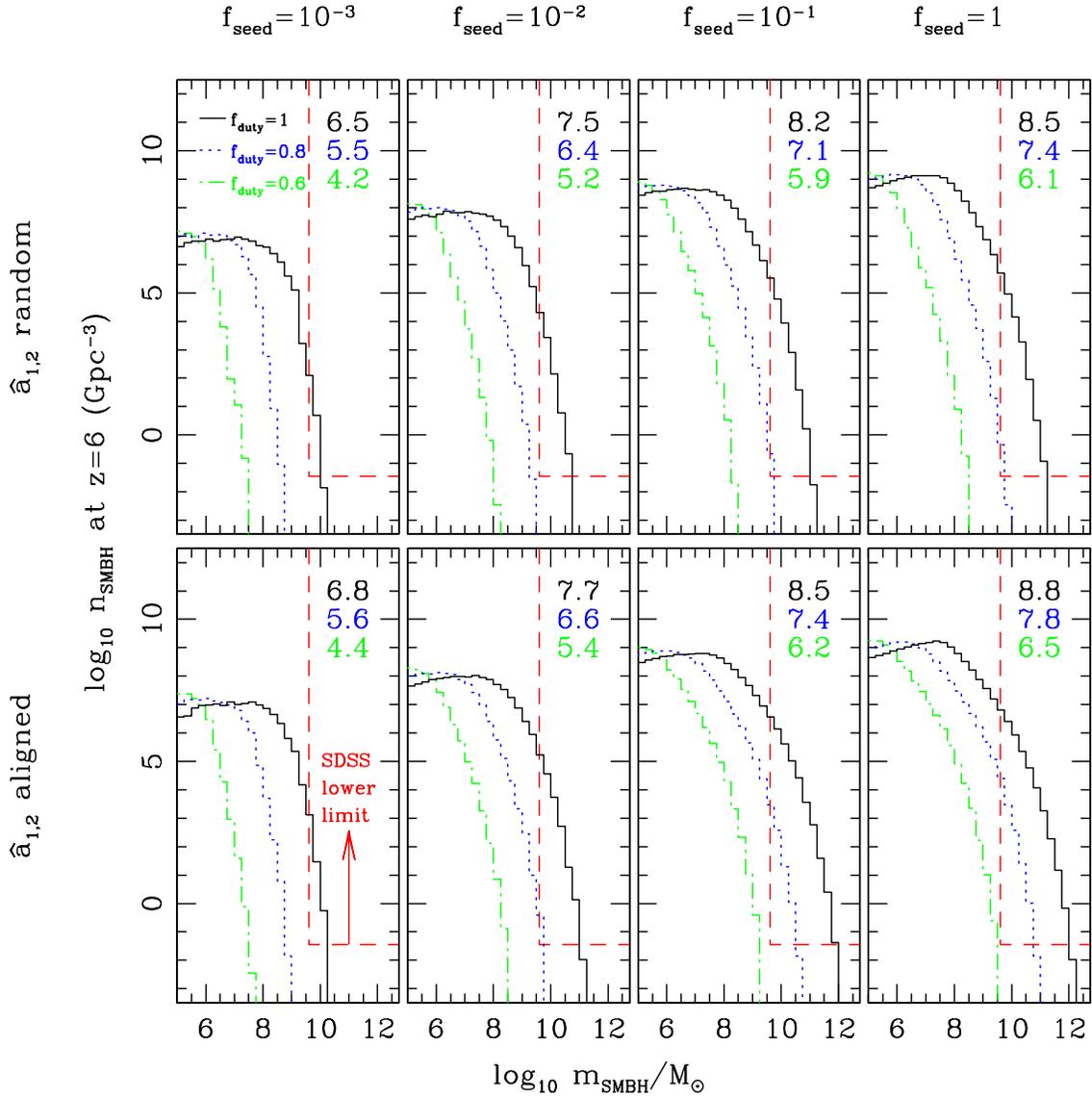}
}}
\caption{The comoving number densities of SMBHs in different
  mass bins at redshift $z=6$.  Colored figures are available in the online version.
  The 24 different models shown in the figure assume different
  parameter combinations as follows. The columns, from left to right,
  adopt $f_{\rm seed}=10^{-3}$, $10^{-2}$, $10^{-1}$, $1$.  The top
  row is for simulations with random binary spin orientation, and the
  bottom row is for spins aligned with the binary's orbital angular
  momentum.  Time--averaged accretion rates are distinguished by
  color: black (solid, $f_{\rm duty}=1$), blue (dot, $f_{\rm duty}=0.8$), and green
  (dash-dot, $f_{\rm duty}=0.6$). The numbers in the upper-right
  corners represent $\log_{10}[\rho_{\bullet}/(\Msol \Mpc^{-3})]$ for
  each model, in descending order of $f_{\rm duty}$.
  The red (dashed) line demarcates the rough indication for the minimum
  number of $z\approx 6$ SMBHs in the observable universe
  with $m\gtrsim 10^{9.6}\Msol$ given the area surveyed by SDSS.   }
  \label{fig:mfpop3}
  \end{figure}

We begin with Figure \ref{fig:mfpop3}, showing the mass function for
the $m_{\rm seed}=100 \Msol$ Pop III seed model.  This and the
companion figures are organized with different values for $ f_{\rm
seed}$ in different columns, the two spin prescriptions in different
rows, and the different duty cycles in different line styles (and
different colors, in the online version).  This will be the default
organization of our 8-panel figures unless noted otherwise.

The numbers in the upper-right-hand corner of each panel represents
the total SMBH mass density, $\log_{10}[\rho_{\bullet}/(\Msol
\Mpc^{-3})]$, for all BHs with $m>10^{5}\Msol$.  Because of
statistical fluctuations, for multiple model realizations with
identical parameters this value can vary by $\ltsim 10\%$ for $f_{\rm
seed}\gtrsim 10^{-1}$, and as much by a factor of a few for $f_{\rm
seed}\ltsim 10^{-3}$.  For each of these simple models, this density
is exceedingly high compared to the SMBH density of the local
universe.  For the present discussion, we will overlook this point as
we address the effects of the various model factors; we will introduce
the additional constraint from $\rho_{\bullet}$ in subsequent
sections.

The most stringent observational requirement for the high-mass end of
the $z\approx 6$ SMBH mass function comes from the SDSS observation of
the $z\approx6.4$ quasar J1054+1024, which has an inferred mass of
$\sim 4\times 10^{9}\Msol$.  Since this object was detected in a
region $\sim 10\%$ of the sky, we estimate that $\gtrsim 10$ similar
objects may exist at $z\sim 6$.  We have adopted this as a rough
indication of the lower limit of the SMBH mass function at redshift
$z=6$, and represent this limit by the upper right quadrangle in each
of the panels, delineated by the red dashed lines.  Note that these
lower limits are conservative, since they do not require the presence
of any additional SMBHs with comparable mass that are ``off''. Since
high values for the duty cycle -- near unity -- are suggested by
quasar clustering measurements (as discussed above), and are also
required for growing the most massive SMBHs (as we find below), this
correction is only of order a factor of $\sim$two. As seen in the
figure, the high--mass end of the SMBH mass function is so steep that
increasing the lower limit on the required SMBH space density by even
$\sim 2$ orders of magnitude would make little difference to our
conclusions.

The first conclusion one can draw from this figure is that several
combinations of parameters can produce SMBHs massive enough to power
the quasar J1054+1024.  We will return to this and other observational
constraints in the next subsection, but let us for now focus on
understanding the effects of our various parameters and their
combinations.  In general, the effect of varying each of the
parameters is relatively straightforward to understand.  Increasing
the accretion rate, increasing the seed occupation fraction, and
aligning the spins all tend to result in more massive BHs.  Note that
the accretion rate in Eddington units, $f_{\rm duty}$, is degenerate
with the accretion efficiency, as $\dot{m}\propto f_{\rm
duty}\times(1-\epsilon)/\epsilon$.  For example, a model with
$\epsilon=0.15$ and $f_{\rm duty}=1.0$ is equivalent to
$\epsilon=0.10$ and $f_{\rm duty}=0.63$.  We have used $\epsilon=0.10$
throughout the results presented here.  With this value for the
efficiency, it is possible to build the most massive SDSS quasar SMBHs
by $z\approx 6$, starting with $100 \Msol$ seeds.  If the efficiency
is $\approx 0.15$, however, it is only possible to build the SMBHs in
question with the most optimistic assumptions: $f_{\rm duty}\approx
1$, $ f_{\rm seed}\gsim 0.1$, and spin alignment would all be
required.

We note two non--trivial observations to be made from Figure
\ref{fig:mfpop3}.  First, if the seed fraction is low, the spin
orientation has a minimal effect on the BH mass function.  This is
because if seeds are extremely rare, they are likely to grow in
isolation for much of their existence along with their host halos.
The first mergers are not likely to occur until the gravitational
potentials of their host halos are deep enough to retain them from any
recoil event.  Second, the increase in the SMBH abundance from
increasing the seeding fraction has a tendency to plateau.  This is
the reverse situation compared to the low $ f_{\rm seed}$ limit: if
seeds are very common, they are likely to experience multiple BH
mergers very early in the merger tree, when their masses are still
comparable, and ejections become very common.  As $ f_{\rm seed}$
increases, assembly becomes increasingly inefficient at early times.
In models where the seeding halo temperatures are lower, the
efficiency saturates at lower values of $ f_{\rm seed}$.  Furthermore,
in models with the shallower TIS gas profiles, we find that increasing
the occupation fraction beyond a certain ``sweet spot'' value rapidly
decreases the final SMBH masses.  The reasons for this are that (i)
the seed BHs in these models barely grow, making near--equal mass
($q\sim 1$) BH mergers, and therefore large kicks, much more common
and (ii) the gas drag is reduced, making it easier to kick holes out
of these halos.  We conclude that arbitrarily increasing the number of
seed holes contributing to the assembly process is not necessarily an
efficient solution to the SMBH assembly problem.  In fact, excessive
seeding can lead to a different conflict with observations --
overproducing the mass density in lower--mass SMBHs --, to which we
will return in the next subsection.

\begin{figure}
\centerline{\hbox{
\plotone{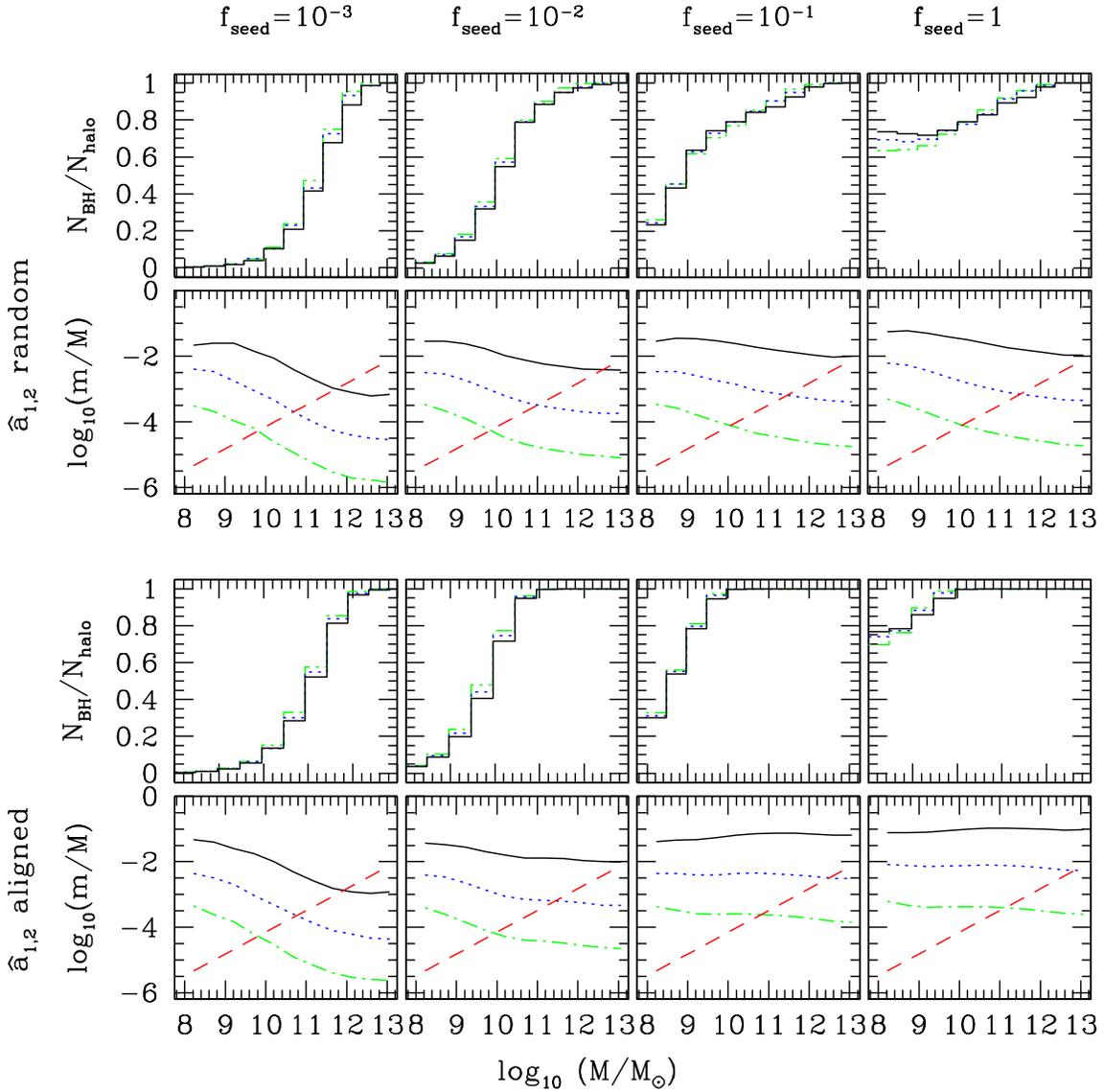}
}}
\caption{Properties of the SMBH population at $z=6$ as a function of
  the halo mass $M$: the percentage of DM halos hosting a central BH
  (assumed at most to be one BH per halo; top rows in both the upper
  and lower panels) with $m\ge10^{-5}M$ , and the mean BH--to--halo
  mass ratio $m/M$ for the halos that do host a BH (bottom rows).
  Color (line-style) and panel schemes are the same as in Figure \ref{fig:mfpop3}.
  The red (dashed) line is the empirical $m/M$ relation extrapolated to
  $z=6$ (see Equation \ref{mM}; Wyithe \& Loeb 2003, Ferrarese 2002).  Our $m/M$
  relation has the opposite trend with respect to halo mass from the
  trend observed in the local universe. Note that in some cases, the 
  central BH consumes most of the baryonic mass in the halo.}
  \label{fig:mmpop3}
  \end{figure}

Let us turn now to Figure \ref{fig:mmpop3}, which shows the BH
occupation fraction in $M>10^{8}\Msol$ DM halos at $z=6$, and the
BH-to-halo mass ratio (which here serves as a proxy for the $m$-$\sigma$
relation; see, e.g. Ferrarese 2002) for each of the models.  Here, we
have arbitrarily defined our occupation fraction to mean the fraction
of DM halos that host a BH with a minimum mass of $10^{4}\Msol$, as we
are interested in the population of nuclear BHs and not stellar
remnants.  Menou et al. (2001) have shown that, in the low-redshift
Universe, the fraction of DM halos hosting a SMBH will approach unity,
regardless of the initial occupation fraction at earlier times.  We
find that at $z\approx 6$, the occupation fraction in halos with
$M\ltsim 10^{11}\Msol$ can still be significantly below unity,
depending primarily on $f_{\rm seed}$. In principle, a future survey
that can determine the quasar luminosity function (LF) at $z=6$ to
several magnitudes deeper than the SDSS could look for this drop in
the occupation fraction, since it will produce a flattening of the LF
at magnitudes below some threshold.  The spin prescription has a
noticeable effect on the $z=6$ occupation fraction for $ f_{\rm
  seed}\gtrsim 10^{-2}$.  On the other hand, the duty cycle
essentially only affects the mass of the BHs, and not their presence
or absence, and so has a minimal effect on the occupation fraction,
within the parameter range shown.  Note that we do not expect to
reproduce the $m$-$\sigma$ relation in our simulations, as we employ
simple and $z$-independent accretion and seeding prescriptions, and
our models have no feedback to enforce the relation.  Instead, the
$m/M$ relation should be taken as a sanity check that we are producing
a physically viable BH population.  Note that in some of the models
shown in Figure 5, the BHs grow much larger than the $m/M$ relation
observed in nearby galaxies. In particular, as the figure reveals,
SMBHs in the lower--mass ($\sim10^8~{\rm M_\odot}$) halos in the
models with $f_{\rm duty}=1$ tend to consume most of the gas in their
parent halos, clearly an improbable result.

For an explicit comparison, alongside the $m/M$ relation produced by
our models, we have plotted the the value expected based on
measurements of the $m$-$\sigma$ relation in local galaxies.
Specifically, we show the expression from Wyithe \& Loeb (2003),
\begin{equation}
\label{mM}
m=10^{12}\epsilon_{0}\left(\frac{M}{10^{12}\Msol}\right)^{\gamma/3}
\left(\Omega_{0}h^{2}\right)^{\gamma/6}(1+z)^{\gamma/2}
\approx 1.16\times 10^{7}\left(\frac{M}{10^{12}\Msol}\right)^{\gamma/3}
(1+z)^{\gamma/2}.
\end{equation}
We adopt their parameter choices of $\epsilon_{0}=10^{-5.4}$ and
$\gamma=5$.  This expression satisfies the SDSS constraints at the
high--mass end of the mass function.  It also agrees well with the
relation suggested by Ferrarese (2002) for SMBHs in the local
universe, $m\sim 10^{7}(M/10^{12}\Msol)^{1.65}$.  As the figure shows,
our predicted $m/M$ relation tends to have the opposite slope than the
one inferred from the observed $m$-$\sigma$ relation: in our results
$m/M$ decreases with mass or stays roughly constant as $M$ increases,
but this is the opposite of the empirical trend.  This is due mainly
to the difference in the growth rates of halos and holes: our simple
prescription for steady, exponential accretion for the BHs can
significantly exceed the growth rate of DM halo masses due to the
accretion of unresolved low--mass halos in the EPS merger tree.  As a
result, in some cases, the host halo mass predicted for the $z=6$
quasars is as low as $10^{11}\Msol$, which is an order of magnitude
lower than would be predicted from the extrapolation of the locally
measured $m/M$ relation, or from the inferred space density of the
host halos (e.g. Haiman \& Loeb 2001).  However, any extrapolation of
the $m-\sigma$ or $m/M$ relation to high redshift, and the masses of
halos that host the brightest $z>6$ quasars, at present, have large
uncertainties, and do not robustly preclude such low values.  As will
be discussed below, the overly rapid growth of SMBHs in this suite of
models motivates modifications to the modeling, including a model in
which the extrapolated $m/M$ relation holds by assumption.

\begin{figure}
\centerline{\hbox{
\plotone{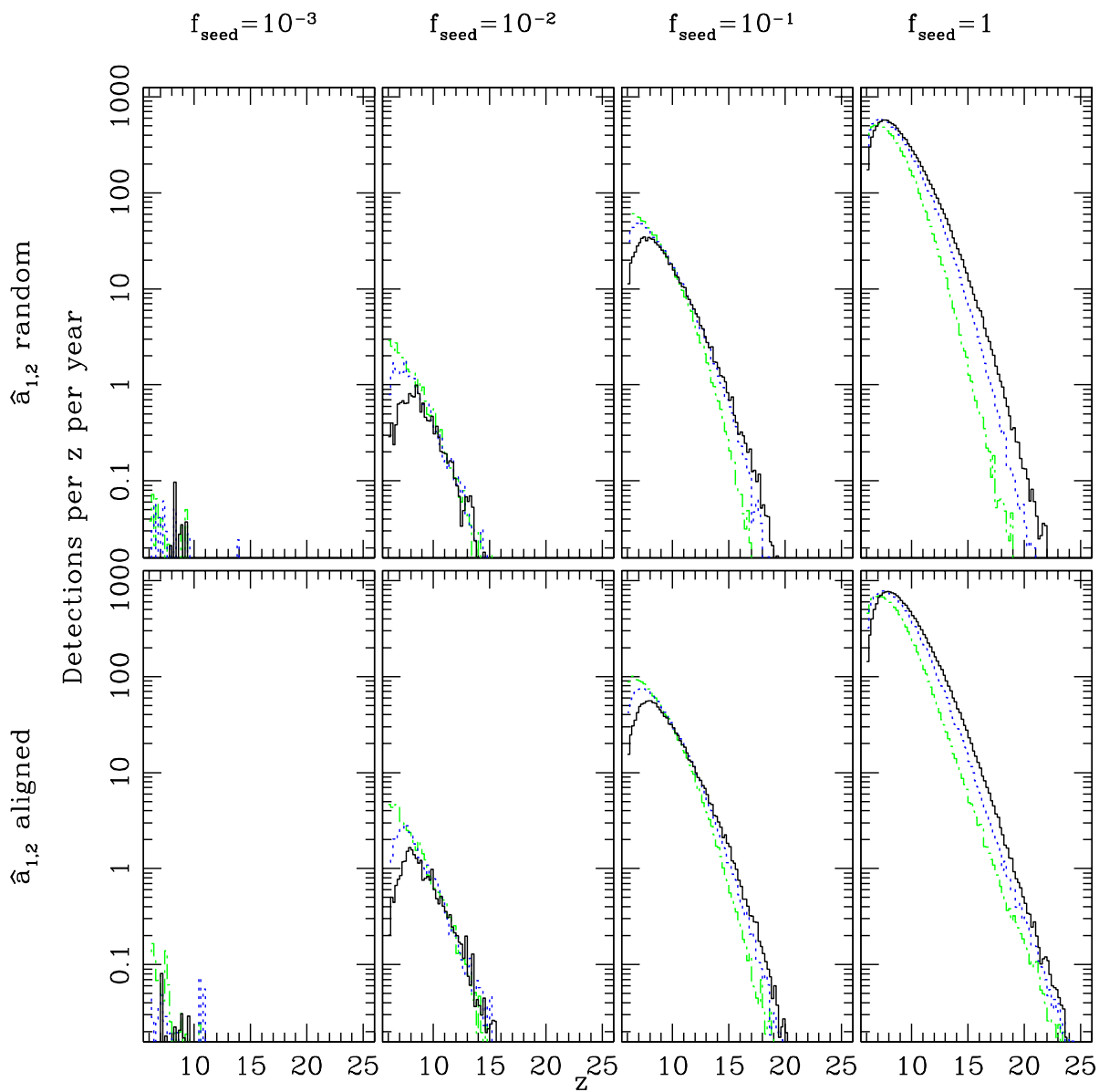}
}}
\caption{Expected number of {\it LISA} detections per redshift per
  year due to SMBH mergers with binary mass $10^{4}\Msol \le
  (m_{1}+m_{2})(1+z)\le10^{7}\Msol$.  The color, line-type and panel schemes are
  the same as in Figures \ref{fig:mfpop3} and \ref{fig:mmpop3}.  }
  \label{fig:LISApop3}
  \end{figure}

Figure \ref{fig:LISApop3} shows estimates for the {\it LISA} event
rate, calculated for all binary mergers satisfying $10^{4}\Msol
<(m_{1}+m_{2})(1+z)<10^{7}\Msol$ as (see, e.g. Menou et al. 2001),
\begin{equation}
\frac{d^{2}N}{dz\; dt}=4\pi c \; d_{\rm com}^{2}(z) \frac{\Delta N}{\Delta z \; \Delta V},
\end{equation}
where $\Delta N$ is the number of SMBH merger events in the tree in a
time step $\Delta z$ and a simulated comoving volume $\Delta V$, and
$d_{\rm com}$ is the comoving distance. Although there is a mild
dependence on the duty cycle / accretion rate and the kick
prescription, it is evident that for our assembly models $f_{\rm
  seed}$ has the greatest effect in setting the rate of SMBH binary
mergers detectable by {\it LISA}. Because the initial merger rates
scale as $f_{\rm seed}^{2}$, the measured event rate is extremely
sensitive to the BH number population.  Since the merger activity
peaks near $z\ltsim10$, {\it LISA} should be able to measure the
masses of most of these SMBH binaries to high prevision (Hughes 2002,
Arun et al. 2008).  Although the raw detection rate -- without any
information on BH spin or mass ratio -- alone will be richly
informative, the ability to determine the source masses with high
fidelity should be very useful in further constraining the growth rate
of the first SMBHs, and breaking degeneracies between the allowed
parameter--combinations.  The observed distribution of binary masses
as a function of redshift will provide direct snapshots of the mass
function and shed light on its evolution, independently from quasar
luminosity surveys (e.g. Yu \& Tremaine 2002; Willott et al. 2006).

\begin{figure}
\centerline{\hbox{
\plotone{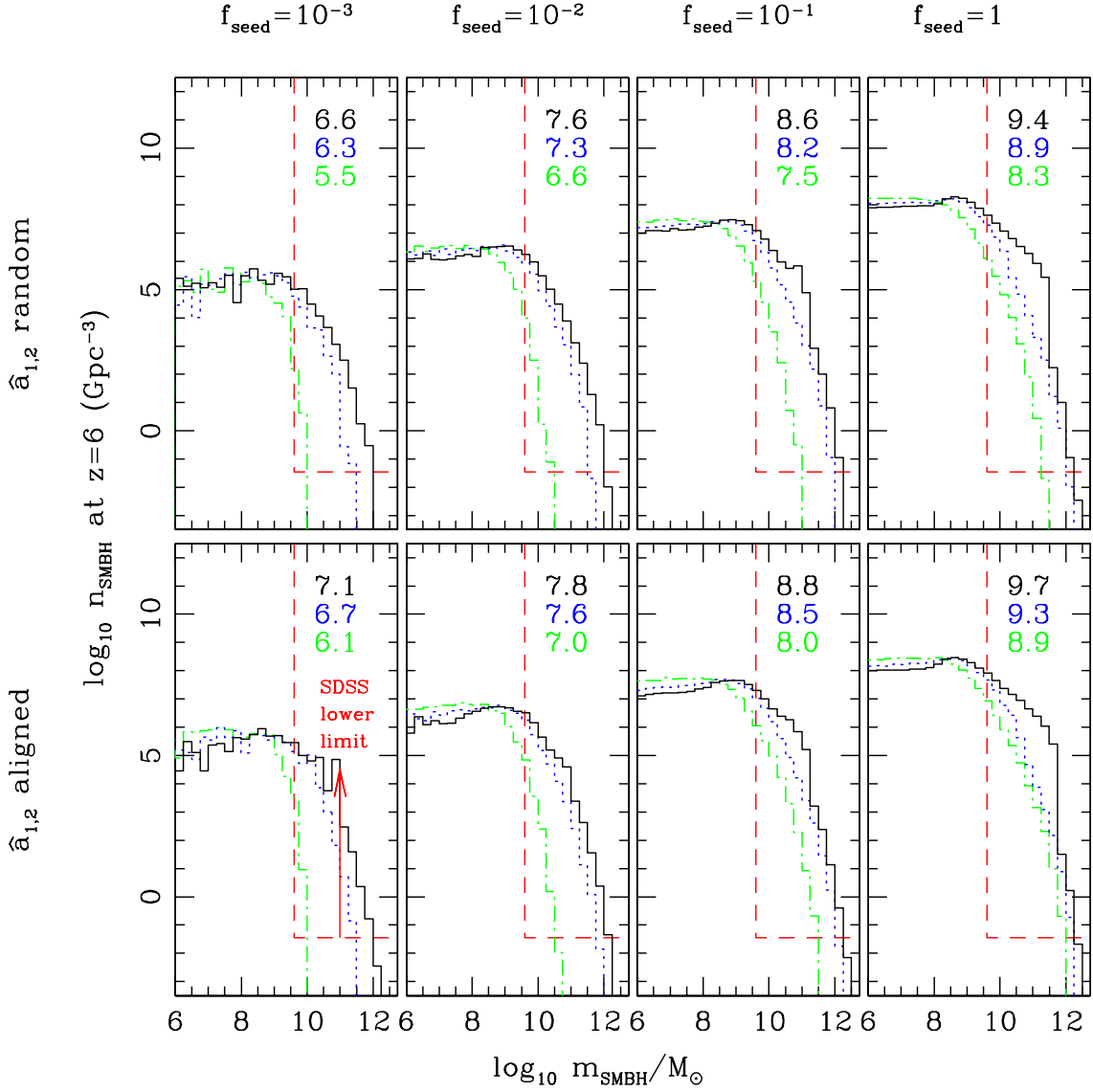}
}}
\caption{The $z=6$ SMBH mass function in the direct collapse
  scenarios, with $m_{\rm seed}=10^{4}\Msol$ and $T_{\rm
    seed}=1.5\times 10^{4}\K$.  Color and panel organization for
  accretion rate, seed fraction and spin alignment is the same as in
  Figure \ref{fig:mfpop3}.}
  \label{fig:mfDC}
  \end{figure}

\begin{figure}
\centerline{\hbox{
\plotone{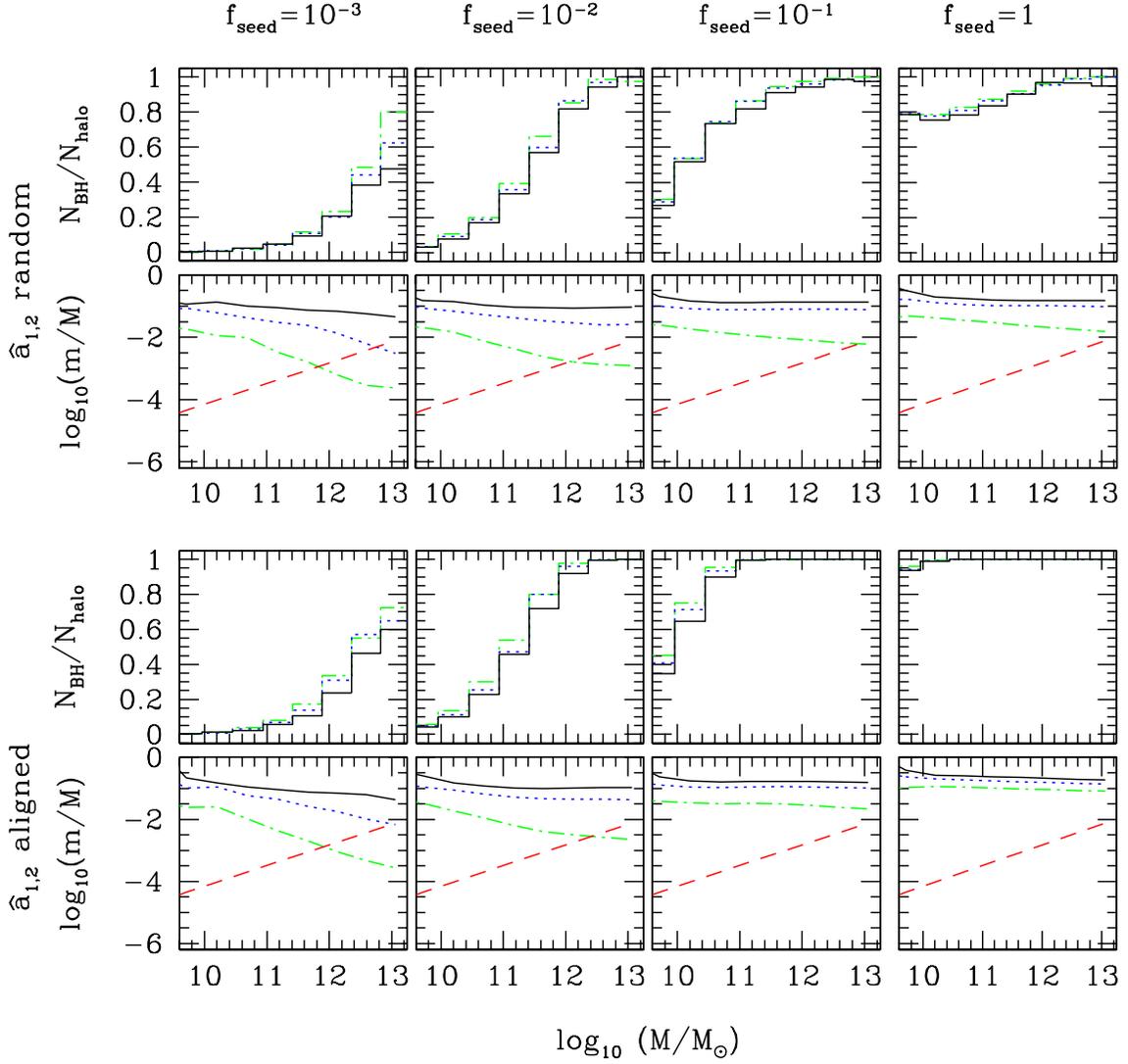}
}}
\caption{The SMBH occupation fraction and the $m/M$ ratios in the
  direct collapse models.  Refer to Figure \ref{fig:mmpop3} for color and
  panel organization.  The dotted line is the extrapolated empirical
  $m/M$ relation.}
  \label{fig:mmDC}
  \end{figure}

\begin{figure}
\centerline{\hbox{
\plotone{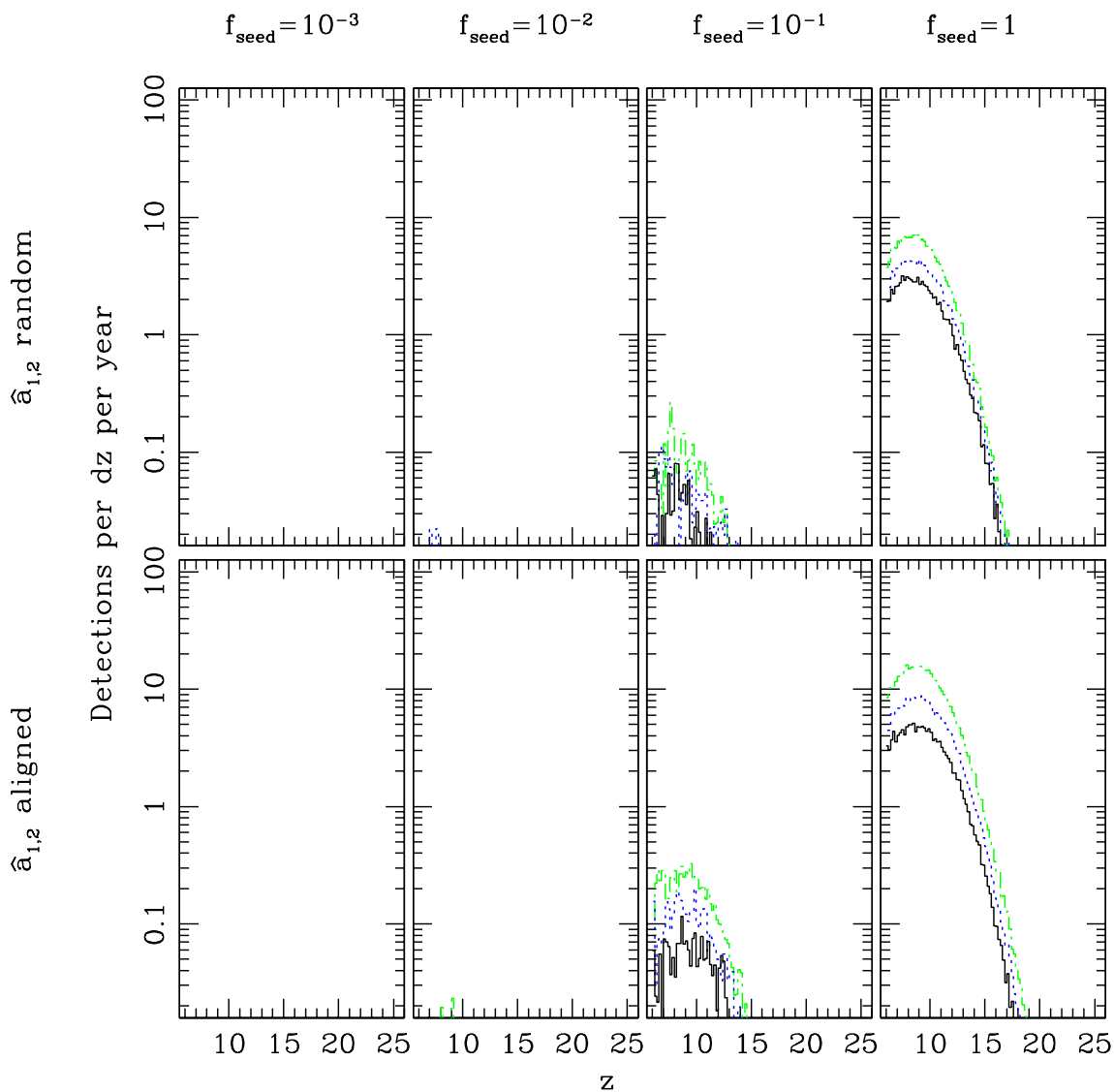}
}}
\caption{ {\it LISA} event rates in the direct collapse scenarios.
  Refer to previous figures for color and panel organization.  Note
  the significant reduction in the event rates relative to the
  pop--III seed models, owing to the smaller number of seed-forming
  halos in the tree.  }
  \label{fig:LISADC}
  \end{figure}

Another family of assembly models that has been frequently discussed
in the literature is the so-called ``direct collapse'' model, wherein
BHs with $m\sim 10^{4-6}\Msol$ are formed rapidly from gas cooling via
atomic H in halos with virial temperature $T\gtrsim 10^{4}\K$ (Oh \&
Haiman 2002; Bromm \& Loeb 2003; Volonteri \& Rees 2005; Begelman et
al. 2006; Spaans \& Silk 2006; Lodato \& Natarajan 2006).  We simulate
such a family of models, for the same set of the parameters $f_{\rm
  seed}$ and $f_{\rm duty}$ and the same spin alignments.  We choose
$m_{\rm seed}=10^{5}\Msol$ and $T_{\rm seed}=1.5\times 10^{4}\K$.  We
show the mass functions, the hole-halo relations and the {\it LISA}
rates (the same information as in Figures \ref{fig:mfpop3},
\ref{fig:mmpop3} and \ref{fig:LISApop3}) in Figures \ref{fig:mfDC},
\ref{fig:mmDC} and \ref{fig:LISADC}.  Although the main differences
are all fairly intuitive, it is instructive to address how the
direct-collapse models differ from the Pop-III seed models.

First, from Figure \ref{fig:mfDC} we see that it is much easier to
construct more massive SMBHs owing to the larger seed masses.  In
fact, the models with $f_{\rm seed}\gsim 0.1$ become unphysical, since
the SMBHs in these models would exceed the baryon mass
$(\Omega_b/\Omega_m)M$ of their parent halos.  The second point is
that the mass function is very differently distributed between the
Pop-III and direct-collapse scenarios.  The reader will note that for
each set of parameters, the overall SMBH density does not differ
significantly between the corresponding seed models.  However, this is
deceiving as the mass function is clearly different, with the
direct-collapse seeds resulting in a more ``top heavy'' SMBH
population.  For the range of parameters surveyed, the Pop-III model
has $\gtrsim 90\%$ of the SMBH mass in the $m<10^{7}\Msol$ range if
$D=0.6$, compared to $\ltsim 30 \%$ for the $D=0.6$ direct-collapse
seed models.  For $D=1.0$ and random spin alignment, $\sim 2-5\%$ of
the SMBH density resides in the most massive $m>10^{9}\Msol$ BHs if
the seeds are Pop III; for the same parameters, $\sim 60-70\%$ of the
mass is in the billion-plus solar mass BHs in the corresponding
direct-collapse models.  Note that the total BH mass density remains
extremely high; again, we will address this point shortly.

Third, there is a slightly weaker dependence on the spin orientation
of the BH binaries.  This is most apparent by comparing the $z=6$
occupation fractions in Figures \ref{fig:mmpop3} and \ref{fig:mmDC},
and is due to the deeper potentials of the host halos in the direct
collapse case.  Fourth, even though the $m/M$ relation continues to
have a slope opposite to the locally observed trend, there appears to
be a break in the relation at $M\sim 10^{9}\Msol$, accompanied by a
drop in the occupation fraction.  This is due to the simple fact that
halos below this mass threshold cannot have many $T>1.5\times
10^{4}\K$ progenitors and the model similarly prohibits
intermediate-mass BHs.  This cutoff contributes to the top-heaviness
of the mass function for these models.  Fifth, {\it LISA} rates are
lower by one to two orders of magnitude than in the Pop-III seed
models, because there are fewer $1.5\times 10^{4}\K$ halos than $1200
\K $ halos (another factor is that the seed BHs are already born with
a mass near the middle of {\it LISA}'s logarithmic mass range, so they
spend only $\sim$ half the time in the {\it LISA} band, compared to
the Pop-III seeds).
It is worth noting, in particular, that it is possible to build the
SDSS quasar BHs in ways that produce no detectable GW events for
observation with {\it LISA} beyond $z\sim 6$ (in contrast, in the
successful pop--III models, a minimum of a few events are predicted).
In such scenarios, SMBHs are extremely rare until $z\sim 6$, at which
point the SMBH occupation fraction will evolve toward unity fairly
rapidly, as described by Menou et al. (2001).

\begin{figure}
\centerline{\hbox{
\plotone{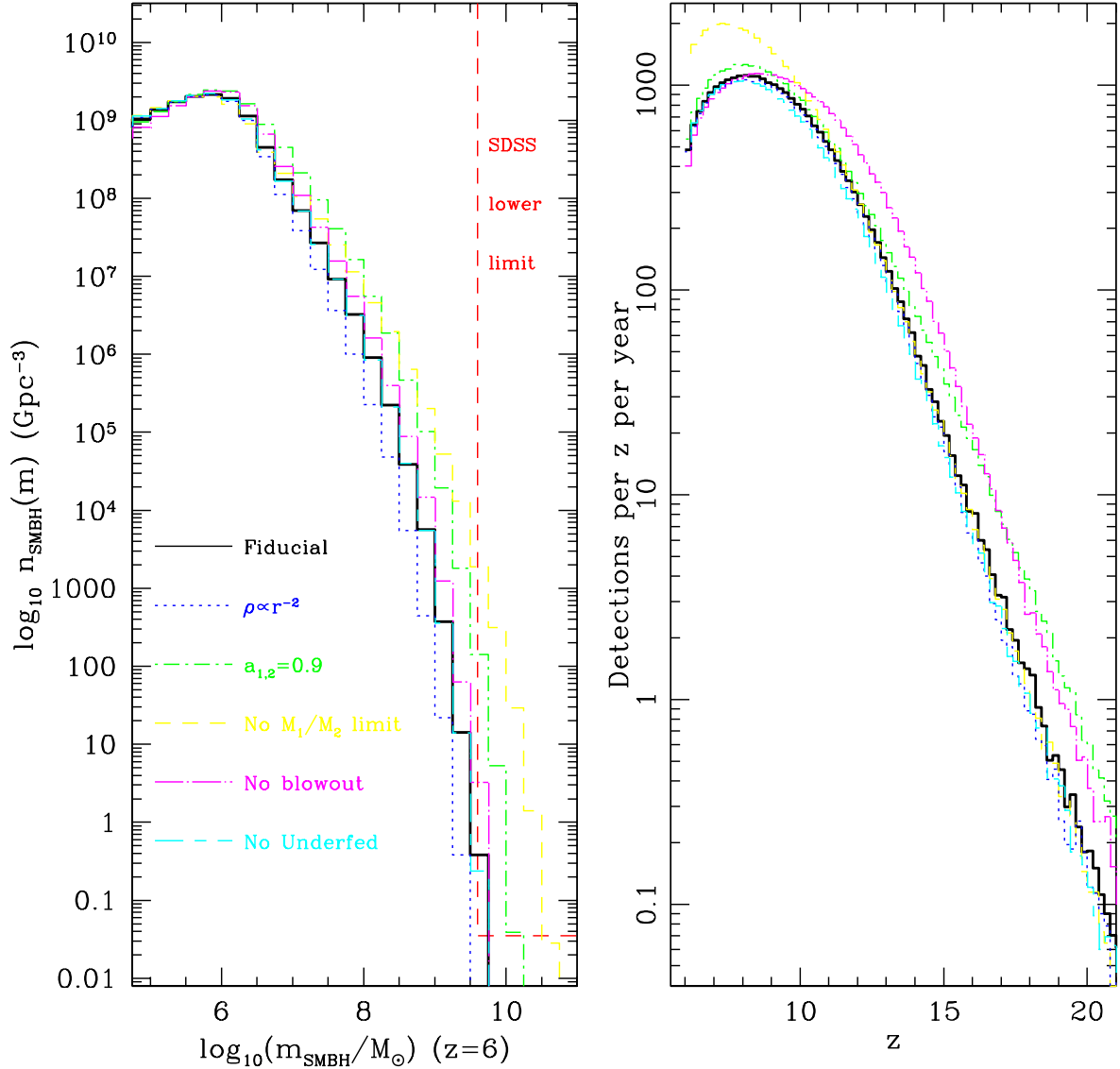}
}}
\caption{Properties of the SMBH population under several variants of
  our fiducial models.  All models plotted have $f_{\rm duty}=0.65$, $
  f_{\rm seed}=1$, and aligned binary spins, and modify a single
  aspect of the basic fiducial model prescription, as labeled: the gas
  density is an isothermal power--law instead of the fiducial
  $\rho\propto r^{-2.2}$ (dark blue, dotted curve); spin magnitudes are
  near-maximal at $a_{1,2}=0.9$ (green, dot-short-dash curve); halos are allowed to
  merge and form BH binaries irrespective of their mass ratio (yellow, dashed
  curve); Pop III stars do not blow out the gas from their host halos
  prior to leaving a seed BH (magenta, dot-long-dash curve); and recoiling BHs wandering
  in low-density regions continue to accrete efficiently (light blue, long-short-dash
  curve).  We also ran simulations with a corey, TIS gas profile for
  the host halos.  We were unable to produce SMBHs of even $\gtrsim
  10^{7}\Msol$ in those models even when prescribing the most
  optimistic values for the other assembly parameters (therefore
  results from these models are not shown in the figure).}
  \label{fig:mfvar}
  \end{figure}

Finally, in Figure \ref{fig:mfvar} we present the mass function and
the {\it LISA} detection rate in five variants of a fiducial $f_{\rm
  duty}=0.65$, $ f_{\rm seed}=1$, aligned spin model with pop-III BH
remnant seeds.  We choose these values because they just barely are
able to match the abundance of the most massive SDSS quasar BHs, and
because they produce the most BH mergers, and thus the effects of
varying the other recoil-related parameters will be the most
discernible. We show the fiducial model in bold.  In the modified
models, we (i) allow the gas density profile to be shallower, with a
$r^{-2}$ power law -- this is to allow for the possibility that the
gas surrounding the BH has not cooled and condensed to high density
(dark blue curves); (ii) require the BH spins to be near-maximal at
$a_{1,2}=0.9$, instead of choosing them randomly from the range $0.0$
to $0.9$ -- this is to allow for the possibility that all SMBHs at
high $z$ are rapidly spinning, e.g. due to coherent accretion (green
curves); (iii) allow halos of all mass ratios to merge, where
previously we had considered a halo with mass less than $1/20$th that
of its merger companion to become a satellite (yellow curves); (iv)
assume that the Pop III seed progenitors are not able to blow out the
gas in the host minihalo (red curves); and (v) ignore the effects of
accretion suppression due to episodes of recoil-induced wandering, and
instead assume that all BHs accrete at $f_{\rm duty}$ unless ejected
(light blue curves).

We also ran models with a TIS profile for the gas component. We found
that these models always failed to produce any SMBHs above
$10^{6}\Msol$ by redshift $z=6$ due to the initial phase of
sub--Eddington accretion of seed BHs, and their results are not shown
in Figure \ref{fig:mfvar}.  This implies that SMBHs must be
continuously surrounded by dense cores of gas that was able to cool at
the centers of DM halos -- feeding holes with the low--density gas in
DM halos whose gas was unable to cool does not allow for high enough
accretion rates (Turner 1991 emphasized the same issue for the growth
of $z\sim 4$ quasar BHs).

The results shown in Figure \ref{fig:mfvar} give insight to the
importance of the assumptions that went into our models.  In
particular, of all the parameters that we have varied for fixed
$f_{\rm seed}$ and $f_{\rm duty}$, the most significant for the SMBH
mass function, by far, are the spin orientation and the limit on the
halo mass ratio for timely mergers.  Both of these have a similar
effect of increasing the number of BHs, especially at the massive end.
The former result -- that maximizing the spin increases the SMBH mass
function -- may seem surprising at first, since generally large spins
imply larger kicks.  However, this is not always the case, as can be
understood from equations (5-8). Under the assumption that both spins
are aligned with the orbital angular momentum, $v_{\parallel}=0$, and
$v_{\perp}\propto (a_1-qa_2)$ is maximized for {\em unequal} spins
(i.e. $a_1=1$ and $a_2=0$ for a typical $q\sim 1$); setting
$a_1=a_2=0.9$ therefore eliminates the largest kicks, and allows more
BHs to be retained.  Comparatively smaller differences are visible in
the mass function for the other model variants.  Figure
\ref{fig:mfvar} also shows that adding in the unequal--mass halo
mergers and increasing the spins affect the {\it LISA} event rates
differently: the former adds new events mostly at $z\lsim 10$, where
unequal--mass mergers are more common, whereas increasing the spin
mostly adds new events at $z\gsim 10$.  Interestingly, ignoring the
blow--out has little impact on the SMBH mass function at $z=6$, but it
does shift the {\it LISA} events to higher redshifts, especially at
$z\gsim 10$.  Figure \ref{fig:mfvar} suggests that the {\it LISA}
event rate can be useful in disentangling these three effects.

Perhaps the most surprising inconsequential variation is ignoring
episodes of reduced accretion due to the BH wandering in low-density
outskirts following recoil.  The reason for this is simply that
lengthy oscillating orbits are relatively rare if the central gas
density is high; the BHs either return quickly, or are ejected (by
assumption).  Recall that our definition for a ``retained'' BH is that
the recoiled hole must return to within 1/10th of the halo's virial
radius within 1/10th of the Hubble time.  For low kick speeds, the BH
does not get very far, because the gas provides both a steep
gravitational potential barrier and a strong dissipative sink.  The
orbit will thus be rapidly damped, with only very brief periods of
underfeeding.  If a BH is kicked hard enough to reach the outskirts of
the halo, there is very little dissipation there, to tug it toward
the center or to further damp its oscillation.  Since the radial
velocity at this point is low, it's likely to remain outside 1/10th of
the virial radius for a significant period.  The bottom line is that
the range of initial kicks that would take the BH to the low--density
outskirts to cause significant underfeeding, while still allowing it
to return quickly enough to be ``retained'' is simply negligibly
small.

Blecha \& Loeb (2008) recently performed a more detailed analysis of
the orbits of recoiling SMBHs that include a multi-component halo mass
distribution and three--dimensional orbits.  They report that recoil
velocities of between $100 \km~\s^{-1}$ and the escape velocity lead
to significant suppression of the accretion rate, with SMBHs accreting
only $\sim 10\%$ of its initial mass over $10^{6}-10^{9}$ years of
wandering through the halo.  We note that (1) in our simulations we
ignore the longest--wandering BHs through our prescribed retention
threshold; (2) typical kick magnitudes for SMBHs are lower than $100
\km~\s^{-1}$ in our simulations, a point we explain below; and (3)
their prescription of the baryon distribution results in a lower
central density than our models, as in that paper they are concerned
with typical galaxies at low redshift, and not with minihalos and
protogalaxies.  We conclude that prolonged periods of wandering and
underfeeding are unlikely to have a significant effect on the mass
function of high--$z$ SMBHs as a whole.  Oscillations may play a more
prominent role in the growth history of SMBHs if large kicks are more
common (and the retained holes tend to be marginally retained), if the
halo gravitational potential is more shallow, or if the effect of
dynamical friction on BH orbits is less than what we have considered
in this paper. Also, halo triaxiality (Blecha \& Loeb 2008) and/or a
clumpy mass distribution (Guedes et al. 2008) could increase the time
that kicked holes take to return to the halo center (or they may never
return).  In principle, this may increase the impact of these
oscillations.  However, in practice, the inner, baryon-dominated
regions of the galaxies are likely close to smooth and spherical. The
BHs that are not ejected in our models typically do not leave these
regions and so will not be subject to large changes in their orbits
from these effects.

The results presented thus far seem to paint a relatively simple set
of relevant parameters for SMBH assembly.  There is the seeding
function $f_{\rm seed}$, which governs the BH merger rate and
therefore to a large extent the {\it LISA} event rate.  The event rate
also depends on the time--averaged accretion rate, parameterized here
by the duty cycle $f_{\rm duty}$, and the initial seed mass $M_{\rm
  seed}$, as these set the evolution of the observable mass spectrum.
The seeding prescription and $f_{\rm duty}$ determine the mass
function almost entirely if $f_{\rm seed}\ll 1$.  If $f_{\rm
  seed}\gtrsim 0.1$, then the spins of the BH binary play a role in
setting the recoil speeds and the subsequent evolution.

Once typical spin and mass parameters of merging SMBHs are determined
by {\it LISA}, either by direct measurement, or perhaps by
extrapolating from detections at lower redshift, combined with the
event rate and what is known about the upper end of the mass function,
this information is likely sufficient to give at least a strong
indication on the typical SMBH growth rate and initial seed mass.

Our simulations above also confirm a result reported by Volonteri \&
Rees (2006), namely that SMBHs form primarily through repeated mergers
of the most massive SMBHs merging with less-massive (SM)BHs.  This is
because the gravitational rocket speeds decrease rapidly as the mass
ratio $q$ decreases.  The first few BHs that ``outweigh'' their
neighbors -- be it through being endowed with more mass at birth,
accreting faster or being fortunate enough to survive the first
mergers -- will be less likely to be ejected from their host halos.
This survival advantage becomes a runaway effect, as each subsequent
merger will result in a lower value of $q$ for the next merger.

\subsection{Constraints on the SMBH Mass Function}

Now that we have a first-glance grasp of the assembly parameters and
their most basic observational characteristics, we can turn to
identifying actual candidate models for the formation of $m\gtrsim
10^{9}\Msol$ SMBHs before $z\approx 6$.
  
The suite of models discussed above has demonstrated that there are
several feasible ways to build the SMBHs. These models have so far
focused only on the number density of $m\gtrsim 10^{9}\Msol$ SMBHs,
and ignored indirect constraints that exist on the mass function at
lower masses.  In particular, the total mass density in SMBHs with
masses in the range $10^{6}\Msol \lsim m \lsim 10^{9}\Msol$ in the
local Universe has been estimated by several authors, who find $\sim
4\times 10^{5}\Msol \Mpc^{-3}$ (to within a factor of $\sim$two; see,
e.g., the recent paper by Shankar et al. 2009a and references therein).
Furthermore, a comparison of the locally observed mass density with
the mass density inferred from accretion by the evolving quasar
population suggests that $\sim 90\%$ of the total local SMBH mass
density is attributable to quasar accretion.  This implies that the
total SMBH mass density increased by a factor of $\sim 10$ between
$z\sim 6$ and the local Universe (Yu \& Tremaine 2002, Haiman et
al. 2004; Shankar et al. 2009a), which then places an indirect
constraint on the SMBH mass function, down to $m\sim10^{5}\Msol$, at
$z=6$.

To be specific, we set the following upper limit on the expected total
$z\approx 6$ SMBH mass density in $m\gsim10^{5}\Msol$ SMBHs:
\begin{equation}
\rho_{\rm SMBH, 5+}(z=6)\sim 0.1\times \rho_{\rm SMBH, 6+}(z\approx 0)\sim 4\times 10^{4}\Msol \Mpc^{-3},
\end{equation}
That is, we assume that the total mass density of SMBHs with mass
$m>10^{5}\Msol$ at $z=6$ is at most 10 percent of the total inferred
mass density of SMBHs with mass $m>10^{6}\Msol$ in the local universe.
The major caveat to making such an expectation is that we assume that
the low--mass end of the BH mass function has grown by a factor of 10,
and that high--redshift quasar luminosity surveys have sufficiently
accounted for selection effects of any SMBHs that may be hidden by
inactivity or by being too dim.

The analysis that follows below is similar to that of Bromley et
al. (2004), who considered the upper limit to the $z\approx 6$ SMBH
mass density in weighing the plausibility of various $z=6$ quasar BH
assembly models.  The main difference is that Bromley et al. (2004) did not
consider the gravitational recoil effect. Adding in this recoil makes
the problem significantly worse. This is because the recoil
necessitates more optimistic assumptions in order to build up the
$\sim 10^{9}\Msol$ SMBHs, which tends to increase, by a large factor,
the predicted number of lower--mass $\sim 10^{6}\Msol$ SMBHs, which
arise later, and whose growth is therefore less sensitive to the
kicks. In other words, the kicks preferentially suppress the abundance
of the most massive SMBHs which arise from the earliest seeds; as a
consequence, we predict steeper SMBH mass functions than the models
considered in Bromley et al. (2004).  We performed a series of assembly
calculations explicitly without recoil, and find that these indeed
produce mass functions with a flatter slope and a higher
normalization, owing to greater SMBH masses at the high--mass end, and
a higher occupation fractions at all masses.  If there are no kicks,
less optimistic assumptions are required to produce the SDSS quasars,
and the overall mass density is lower in no-kick scenarios that
produce the minimum number of $\gtrsim 10^{9}\Msol$ SMBHs.

The basic result we find is that among the models presented thus far,
{\it all of those that match the SDSS abundance of the most massive
  SMBHs overshoot the above value by two or more orders of magnitude.}
One possible solution to this apparent problem is that there is no
problem at all: we have simply set our constraint too severely.
Perhaps not all $10^{5}\Msol$ SMBHs at $z\approx 6$ have evolved to
become $m>10^{6}\Msol$ BHs in galactic nuclei in the local universe.
Still, there is no obvious way to ``hide'' these low--mass SMBHs
locally, and the over--prediction in the SMBH densities in our
simulations are very large: we find that in our fiducial models, we
typically need to reduce the population of SMBHs in the mass range
$10^{5-7}\Msol$ by a factor of $\gsim 100$, and those in the
$10^{7-9}\Msol$range by a factor of $\gsim 10$.

\subsection{Successful Models I: BH Seeds Stop Forming Early}

One logical solution, and one that has been suggested by Bromley et
al. (2004), is to simply stop making seeds below some cutoff redshift.
The earliest seeds contribute the majority of the total mass of the
most massive SMBHs, and late-forming seeds tend to end up in
lower-mass SMBHs.  In principle, then, by introducing a cutoff
redshift for seed formation, one can suppress the low-mass end of the
mass function, while still allowing for the formation of the most
massive SMBHs.  Such a model would also be in line with our physical
understanding of seed BHs.  We know Pop III stars stopped forming
relatively early on in the universe, with the halt in production being
due to trace metal contamination (e.g. Omukai et al. 2008), radiative
feedback from the UV and/or X-ray background during the early stages
of reionization (e.g. Haiman et al. 2000), the higher turbulence of
gas in the centers of later halos, or a combination of these factors.

\begin{figure}
\centerline{\hbox{
\plotone{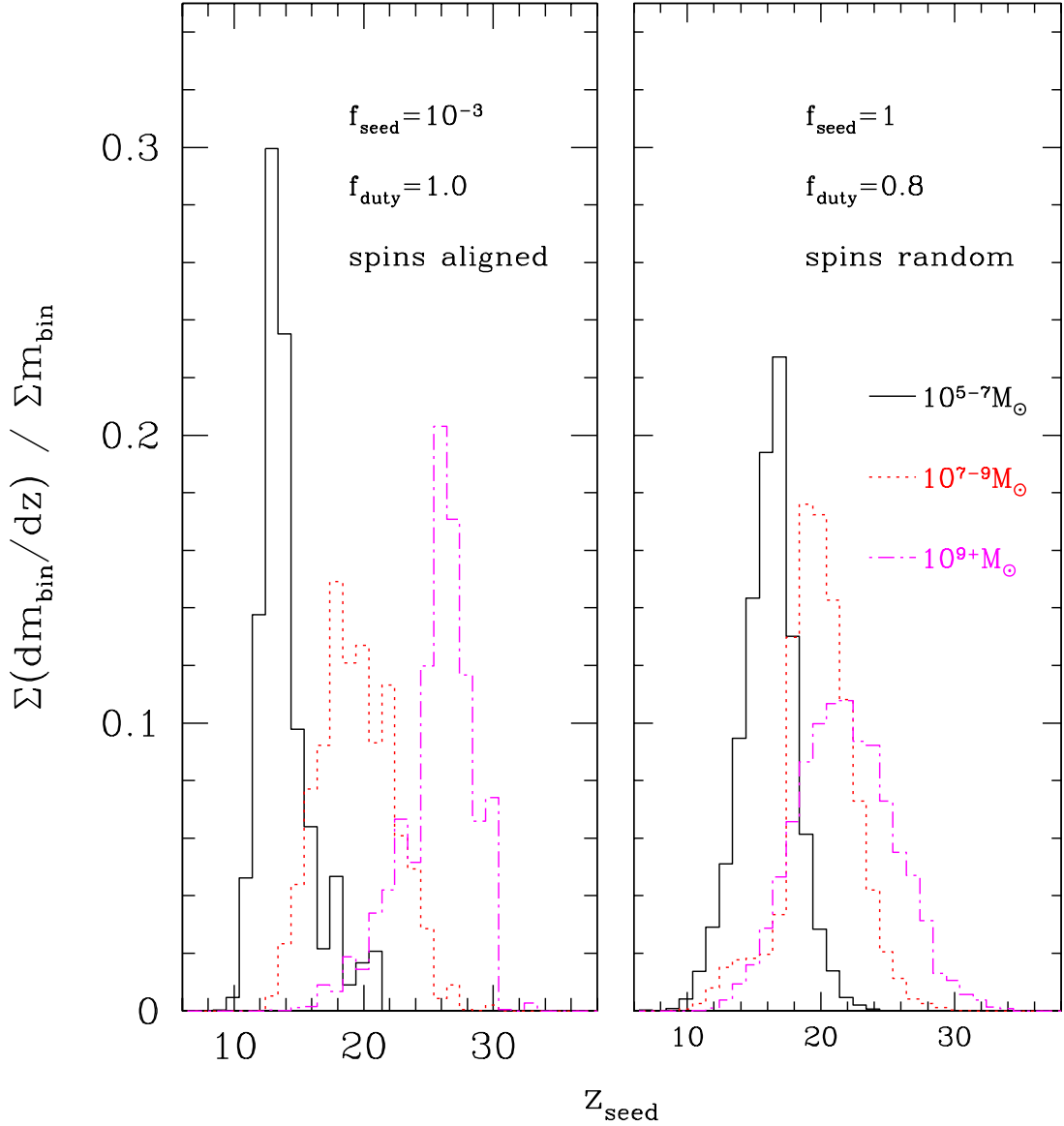}
}}
\caption{The fractional contribution to the mass of $z=6$ SMBHs from
  $100 \Msol$ seed BHs that form at different redshifts $z_{\rm
    seed}$.  The contributions are computed in three mass bins of
  $z=6$ SMBHs: $10^{5}\Msol\le m<10^{7}\Msol$ (black, solid lines), $10^{7}\Msol\le
  m<10^{9}\Msol$ (red, dotted), and $m\ge 10^{9}\Msol$ (magenta, dash-dot).  The most massive $z=6$
  SMBHs arise mainly from the earliest $1200 \K$ progenitors of the
  most massive halos ($z\gsim 20$), whereas the seeds contributing
  most of the mass of lower--mass SMBHs formed later ($z\lsim 20$).}
  \label{fig:frac}
  \end{figure}

This proposed solution amounts to keeping $f_{\rm duty}$ a constant
while allowing $f_{\rm seed}$ to evolve with $z$; in the simplest
case, as a step function dropping to $f_{\rm seed}=0$ below some
redshift.  Figure \ref{fig:frac} shows the fractional contribution
$(dM/dz_{\rm seed})/M$ from seeds forming at different redshifts to
the final mass at $z=6$ in three different bins of the $z=6$ SMBH mass
function, for two different models that are marginally able to
assemble the SMBHs powering J1054+1024.  The model in the left panel
has $f_{\rm seed}=10^{-3}$, while the model on the right has $f_{\rm
  seed}=1$.  Note that contributions to each mass bin are normalized
to integrate to unity, but the two lower mass bins $\rho
(10^{5-7}\Msol)$ and $\rho (10^{7-9}\Msol)$ make up the vast majority
of the total mass density.  The formation epochs of the seeds
contributing the majority of the mass of the different SMBHs are very
distinct in the model with lower $f_{\rm seed}$, but overlap
significantly for $f_{\rm seed}=1$.  What accounts for this
qualitative difference in the assembly epochs?  There are two ways to
build $10^9 \Msol$ holes: accretion onto the earliest-forming holes
that undergo few mergers and grow mostly in isolation, and the
mashed-together product of many seeds that may have formed later.  If
the occupation fraction is high, one expects both populations to
contribute, and therefore a wide distribution for $dM/dz_{\rm
  seed}$. If the occupation fraction is low, SMBHs with many
progenitors contributing become rare, and we find that the most
massive SMBHs can only be formed from isolated seeds in the earliest
minihalos, through few mergers, and so the $dM/dz$ curves in these
models are sharply peaked.

Our task is to eliminate $\approx 99\%$ of the BH mass in the
lower--mass bins, while leaving most of the $m\gtrsim 10^{9}\Msol$
holes unaffected.  By simply examining the normalized $dM/dz$ curves
in Figure \ref{fig:frac}, one can see that simply cutting off seed
formation at an arbitrary redshift for the $f_{\rm seed}=1$ model is
not a viable solution to the overproduction problem, as there is no
way to eliminate the lower-mass SMBHs without eliminating a
significant fraction of the $10^{9}\Msol$ holes.  Cutting off the seed
production can produce successful mass functions for models with
$f_{\rm seed}\ll 1$, but only if the seed cutoff occurs at very high
redshifts, typically $z\sim 30$.  Essentially, the solution calls for
a very brief and early period of seed BH formation, and very rare seed
formation there after.  An unfortunate generic consequence of this
early cutoff redshift is that it quickly chokes off the {\it LISA} event
rates.

We also simulated models where seed production continues beyond the
cutoff redshift -- with the same probability $f_{\rm seed}$ as in the
minihalos at $z>z_{\rm cut}$ -- in halos with $T_{\rm vir}>10^{5}\K$.
Such halos could continue to form BHs if heating by the UV background
is the primary mechanism for seed suppression, as they are able to
shield their central gas from the UV radiation (Dijkstra et al. 2004).
We find that models with such a partial cutoff overproduce the SMBH
mass density if $f_{\rm seed}\gtrsim 10^{-4}$.  This result implies
that either the initial occupation fraction is very low (it is still
possible to make the SDSS BHs with this low $f_{\rm seed}$; see
Table~2 below), or else some other feedback beyond UV radiation, such
as metal--enrichment, stops seed BHs from forming in all halos at
$z<z_{\rm cut}$, even in the rare, more massive ones.

In short, the requirement in models in which the duty cycle is a
constant, but seeds stop forming suddenly below some redshift, can be
simply summarized as follows: the only way to build SMBH mass
functions that satisfy observational constraints and indications at
both the low-mass and the high-mass ends is from extremely rare seeds
that form during a brief and very early epoch.  We also note that
these models are attractive because (i) there are physical reasons for
the seeds to stop forming below some redshift, and (ii) there is
independent empirical evidence, from constraints on the reionization
history from {\it WMAP} measurements of the cosmic microwave
background polarization anisotropies, that the ionizing luminosity in
high--redshift minihalos was suppressed by a factor of $\gsim 10$
(Haiman \& Bryan 2006).

We present in Table 2 the parameters for four such successful models.
While it is not computationally feasible to search the entire
parameter space for such models, we present two typical examples for
both the Pop--III--remnant and direct--collapse seed BH scenarios.
$f_{\rm seed}$ is low in each of these examples, as we have argued
above that they must be.  Although we have listed the spin
prescriptions, they are relatively unimportant because seeds are rare
(see Section 3.1 above).  Given a particular value for $f_{\rm seed}$,
the only free parameters are the accretion rate $f_{\rm duty}$ and the
seed cutoff redshift $z_{\rm cut}$.  For the Pop III models, we find
in the range $f_{\rm seed}\ge 10^{-4}$ that seeds must typically stop
forming at $z_{\rm cut}>30$, with a lower cutoff $z_{\rm cut}\sim20$
for the direct--collapse models.  An important conclusion is that in
each of these models, GW events are too rare ($<10^{-3}/dz/yr$ for
$z<30$) for {\it LISA} detections beyond $z\sim 6$ to occur within the
mission's lifetime.

The mass density of {\it ejected} BHs is exceedingly low in each of
the successful scenarios, $<10^{-3}\Msol \Mpc^{-3}$.  Because the
total number of seeds is small, so are the number of ejected holes.
We only give an upper limit here, because the ejected holes are too
rare for us to give a robust value given the statistical limitations
of our ``halo cloning'' method for populating the entire halo
population at $z=6$ (the mass density in ejected BHs can be large
in models with large $f_{\rm seed}$; see below).

These models represent the simplest scenarios for SMBH formation,
requiring a very brief period of seed formation and a prolonged period
of accretion at rates comparable to the Eddington rate, and
consequentially they represent the most pessimistic predictions for
{\it LISA}'s observational prospects.

\subsection{Successful Models II: Feedback Adjusted to Maintain $m$-$\sigma$ Relation}

While the suppression of BH seed formation is an attractive
possibility that fits constraints on the $z=6$ SMBH mass function, it
is clearly not unique.  One alternative solution to the
over-production problem is to simply reduce the accretion rate of
lower--mass BHs at lower redshifts.  This is again physically
plausible: accretion could be choked off as a result of the baryonic
gas being churned into stars, being heated and dispersed by
reionization feedback, or through self-induced negative feedback where
the BH's own accretion-powered radiation stops the gas supply.  Rather
than try to model such a time-- (and probably mass--) dependent mass
accretion scenario, we examined several model variants, in which BHs
are allowed to accrete just enough mass to match the value inferred by
the $m$-$\sigma$ relation between BH mass and host halo velocity
dispersion.  That is, at each timestep $t\rightarrow t+\Delta t$, all
BHs were assumed to grow in mass by $m\rightarrow m+\Delta m$ such
that the $m-M$ relation is satisfied at the new host halo mass and
redshift.  However, whenever this requires super-Eddington growth,
i.e. if $(m+\Delta m)/m > \exp(\Delta t/t_{\rm Edd})$, then Eddington
growth is applied instead.  The main additional assumption here is
that the $m$-$\sigma$ relation remains valid at all redshifts (which is
at least consistent with a comparison between the evolution of quasars
and early--type galaxies at $0<z\lsim 6$; Haiman et al. 2007; Shankar
et al. 2009b).  As above, we adopt Equation (\ref{mM}) as our
extrapolated $m/M$ relation.

The BHs in these models form as $100\Msol$ seeds, and, given their
host halo mass and redshift, accrete to match this relation as closely
as possible without exceeding the Eddington accretion rate.  If the
simulation completes with the mean BH accretion rate well below the
Eddington rate at all redshifts, then it is consistent with satisfying
the Eddington accretion rate and the $m-M$ relation inferred by
Equation (\ref{mM}).  As we shall find below, in our models the
maintenance of the $m$-$\sigma$ relation does not typically require that
the Eddington accretion rate is saturated (see Figure \ref{fig:accms}
below) as long as the $f_{\rm seed}\gtrsim 10^{-2}$.  If both the seed
mass and the seeding fraction are low, it is increasingly difficult to
satisfy Equation (\ref{mM}) at higher redshifts while simultaneously
satisfying the Eddington upper limit.  We find that for $f_{\rm
  seed}\ltsim 10^{-2}$, accretion must saturate at the Eddington rate
for much of $z\gtrsim 15$ until the extrapolated $m$-$\sigma$ relation
is satisfied, with the mass function falling below this relation at
earlier stages of growth.

Recoil velocities are calculated with the spin magnitudes chosen
uniformly between 0.0 and 0.9.  As with our previous models, we run
simulations where the spins are either randomly oriented or completely
aligned with the angular momentum vector of the binary orbit.  This
class of models in effect represents the most optimistic {\it LISA}
expectations, as it allows us to keep numerous seeds, while simply
adjusting the accretion rate, as described above, to keep the mass
function within bounds.  Note that the recoil speeds are also
minimized by our choice for the spin alignment.

\begin{figure}
\centerline{\hbox{
\plotone{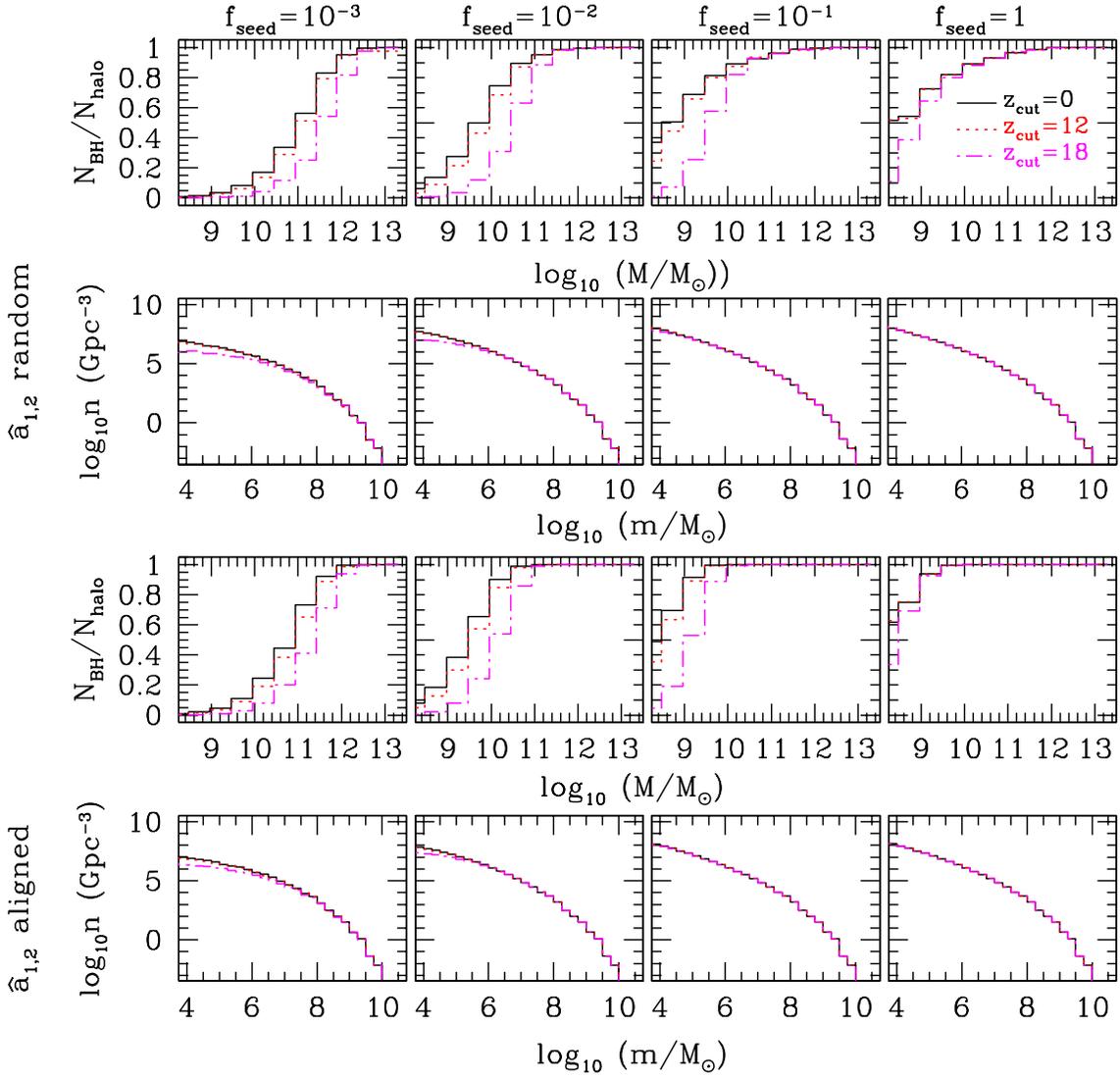}
}}
\caption{ 
The SMBH mass functions and occupation fractions at $z=6$ for various models satisfying the $m$-$\sigma$ relation.
These mass functions are much less steep in comparison to those
shown in Figures \ref{fig:mfpop3} and \ref{fig:mfDC}.  Occupation fractions are higher than those
in Figures \ref{fig:mmpop3} and \ref{fig:mmDC}, as the BH binary mass ratios are generally lower
compared to those models, resulting in less powerful recoil kicks.
}
  \label{fig:mfms}
  \end{figure}

We show the mass functions and occupation fractions for this model in
Figure \ref{fig:mfms}, for three different values of the cutoff
redshift below which new seeds are not formed, $z_{\rm cut}=0$, $12$
and $18$.  Note that it is still possible to form the SDSS
quasar-SMBHs via Eddington-limited accretion by $z\approx 6$ even if
seeds form in just $0.1\%$ of all $1200\K$ halos and only before
$z=18$.  We do not plot the $m/M$ relation, as it is satisfied in the
form of Equation (\ref{mM}) in all cases shown here, by construction.
These models also satisfy, by construction, the upper limit for the
SMBH mass density.  In all of the models shown in Figure
\ref{fig:mfms}, $\rho_{\rm SMBH, 5+}(z=6)\gtrsim 1.3\times10^{4}\Msol \Mpc^{-3}$ if $f_{\rm seed}\ge 10^{-2}$.  In the $f_{\rm
  seed}=10^{-3}$ models, we find $\rho_{\rm SMBH, 5+}(z=6)\sim 1.0\times
10^{4}\Msol \Mpc^{-3}$.  The difference in $\rho_{SMBH}$ is due to
varying occupation fractions at the low end of the halo mass function.
The mass functions in Figure \ref{fig:mfms} have a much shallower
slope overall than those in Figures \ref{fig:mfpop3} and
\ref{fig:mfDC}.  For the mass functions shown earlier, the steeper
slopes were due to the ratio $m/M$ increasing with decreasing host
halo mass; the observed $m/M$ relation has the opposite trend.
  
\begin{figure}
\centerline{\hbox{
\plotone{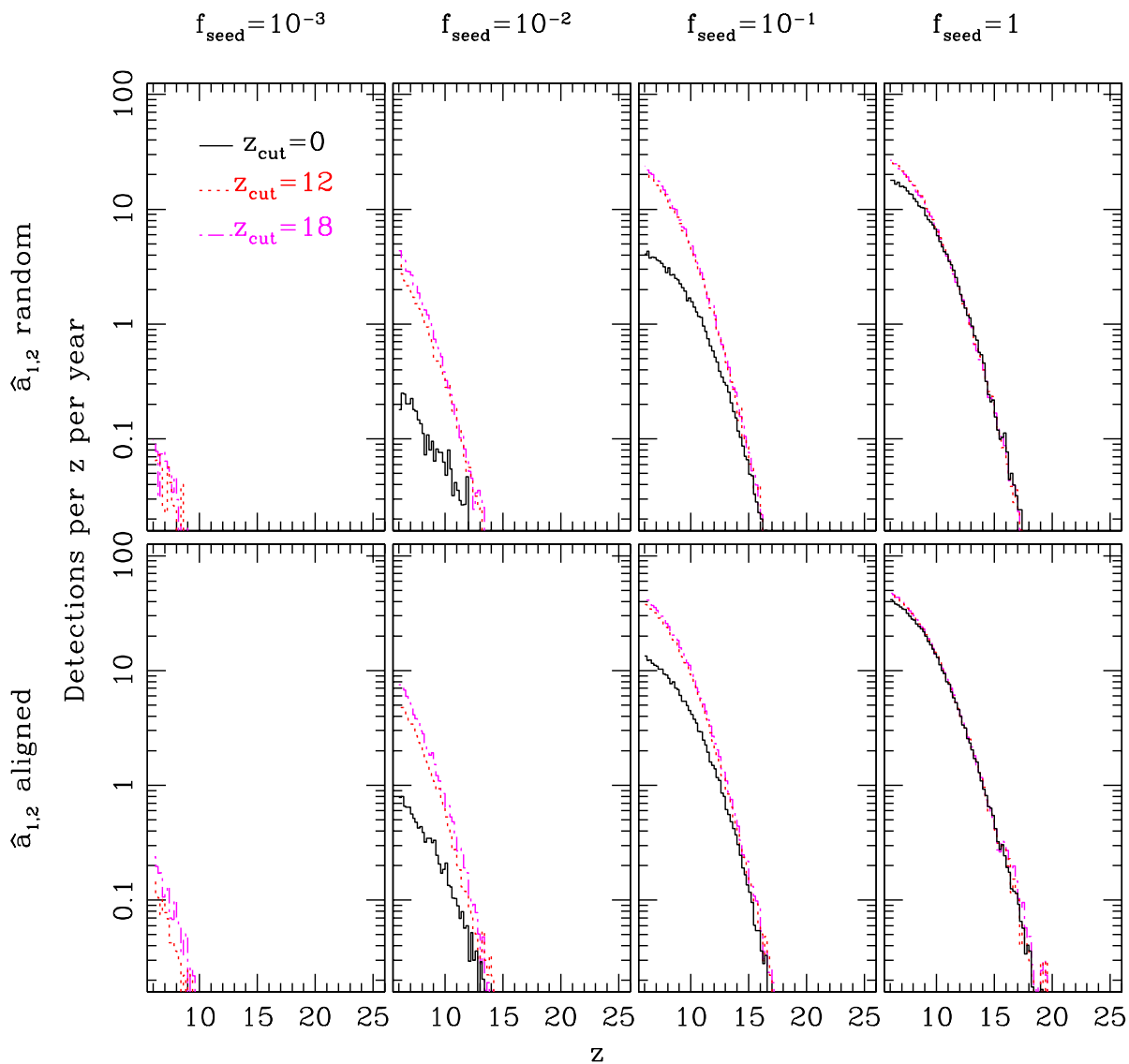}
}}
\caption{The {\it LISA} rates in models that satisfy the $m$-$\sigma$
  relation at all redshifts as closely as possible without exceeding
  the Eddington accretion rate.  Note that the rate saturates at $30 \yr ^{-1}$
  at $z\gtrsim 6$ for
  high seed fractions and low redshifts for seed cutoff.}
  \label{fig:LISAms}
  \end{figure}

We show the {\it LISA} event rates in these alternative models in
Figure \ref{fig:LISAms}.  Most significantly, we note that in the
$f_{\rm seed}=1$ versions of these successful models, the {\it LISA}
event rate can be as high as 30 yr$^{-1}$.  (Note that this number can
be even higher if seeds can form in minihalos down to a virial
temperature that is significantly lower than our assumed fiducial
value of $1200 \K$.)  The rate is somewhat suppressed when compared to
the earlier Pop III seed models (Figure \ref{fig:LISApop3}) that
exceeded realistic indications on the SMBH mass density, because the
massive BHs in the {\it LISA} band are rarer due to the more modest
growth rates.  We draw the attention of the reader to the apparent
independence of the detection rate on the seed fraction and seed
formation cutoff in the cases where $f_{\rm seed}\gtrsim 10^{-1}$ and
$z_{\rm cut}\ltsim 12$.  Because BH ejections probabilities are lower
in these models when compared to the constant-accretion scenarios of
Figures \ref{fig:LISApop3} and \ref{fig:LISADC}, and because the $m/M$
ratios are the same function of $M$ in all models shown, the {\it
LISA} rates saturate and converge once the occupation fraction in the
most massive halos approach unity.  Because the $T=1200\K$ halos form
in greatest abundance at $z\sim 20$, the $z_{\rm cut}=12$ case hardly
differs from the case with no seed cutoff.

A key characteristic of any SMBH assembly scenario is the balance
between growth through BH mergers and growth through gas accretion.
As discussed above, the two must strike a balance such that they are
able to account for the most massive observed quasar-SMBHs at $z\sim
6$, while also not exceeding the total observed SMBH mass density.  In
Table 3, we illustrate the relative importance of mergers {\it vs.}
growth in the models presented in this paper: the four successful
constant-accretion models from Table 2; two of the unsuccessful
constant-accretion models, also in Table 2, which overproduce the
universal SMBH mass density; and four of the models that explicitly
follow the extrapolated $m$-$\sigma$ relation via Equation (\ref{mM}).
The values shown in the table (in $\log_{10}$) are the sum of the
initial masses of all the seed BHs that enter our merger trees; the
total mass of galactic BHs at $z=6$
\footnote{Note that the values in Table 3 include BHs of all masses
equal to and above the seed mass, where we have considered only those
with $m\ge 10^{5}\Msol$ in computing the universal ``SMBH'' mass
density in previous sections.  Also note that we do not generate trees
for DM halos with $M(z=6)<10^{8}\Msol$, and throughout this paper we
do not account for any BHs that may reside in such halos.}; and the
total mass of BHs ejected before $z=6$.  We also calculate the ratio
of the total (galactic and ejected) BH mass at $z=6$ to the total
initial seed BH mass, which gives a simple measure of the growth
through gas accretion.  We see immediately the contrast between the
two types of successful models: the constant-accretion scenario relies
on gas accretion for much of the growth, typically several orders of
magnitude in the total BH mass; where the self-regulating models
essentially describe the most heavily merger-driven scenarios
possible, requiring accretion-driven growth of as little as a factor
of a few.

These models also produce a significant population of ejected BHs.
Even though ejection rates are lower on the whole than our
constant-accretion models (compared to the unsuccessful constant
accretion Model X in Tables 2 and 3), two factors contribute to the
ejected BH mass being comparable to the galactic BH mass at $z=6$.
First, seed BHs are allowed to be very common, especially in contrast
to the successful constant-accretion Models A through D; this results
in a far greater number of total merger events, and a high total
number of ejections despite the lower ejection probabilities.  Second,
the surviving BHs do not grow nearly as rapidly in these models as in
the constant-accretion scenarios, so that the ratio between the total
galactic BH mass and the total ejected mass can remain low (whereas in
the constant accretion scenarios, retained holes can rapidly outgrow
their escaped counterparts.)  In the $f_{\rm seed}=1$ models, the
ejected holes can outnumber and outweigh their retained galactic
counterparts with mass densities of $\sim 3\times 10^{4}\Msol
\Mpc^{-3}$ and number densities of $\sim 100 \Mpc^{-3}$.  The mass
function of the ejected holes is peaked slightly above the seed mass
(because holes are most likely to be ejected at the earliest stages of
their evolution, when their host halos are the least massive).  All of
the ejected BHs are $\ltsim 10^{4}\Msol$ if spins are aligned, but in
rare instances, SMBHs as massive as $\sim 10^{8}\Msol$ are ejected in
our models with randomly oriented spins (the ejected SMBHs with masses
above $m>10^{6}\Msol$ have a very low number density, SMBHs of
$\sim4\times10^{-5}\Mpc^{-3}$, even in the model with the most
ejections (random orientation, no cutoff redshift, $f_{\rm seed}=1$).

These self-regulating accretion models work by adjusting the BH
accretion rates according to the mass growth of their host halos.  We
plot the accretion rates in units of the Eddington rate for this new
set of models in Figure \ref{fig:accms}.  We do so for the $z_{\rm
cut}=12$ case with four combinations for $f_{\rm seed}$ and spin
alignment, and for three different BH mass ranges: $10^{3}\Msol\le m
\le10^{6}\Msol$, $10^{6}\Msol\le m \le10^{8}\Msol$, and
$m\ge10^{8}\Msol$.  Note that the accretion rates must be slightly
higher if BH binary spins are randomly oriented, in order to
compensate for the higher ejection rates.  Similarly, accretion rates
are higher if seeds are less common, in order to compensate for the
reduced merger-driven growth.  For the models shown, the duty cycle
(the mean accretion rate in Eddington units) for the most massive
SMBHs converge to $\sim 0.2$ at $z\approx 6$, though it can be as high
as $\gtrsim 0.5$ at $z\gtrsim 8$ if merger-driven growth is hindered
by low occupation fraction or recoil-induced ejections.

We note that similar merger--tree models tracking SMBH growth have
been published.  For example, Koushiappas et al. (2004) presented a
similar model where the SMBHs are assembled primarily through mergers
of directly-collapsed halo cores.  Bromley et al. (2004) also
presented SMBH assembly model wherein gas accretion activity was
triggered by major mergers of the BHs' host halos; in their model, a
set fraction of the baryonic mass of the host was fed to the BH at
each major merger.  Their prescription (albeit without gravitational
recoil) successfully produced the most massive SDSS SMBHs before
redshift $z=6$ without overproducing the mass density.  In general,
this type of assembly model is fairly easily tuned to broadly
reproduce the $m-\sigma$ relation, as the parallel mass growth of BHs
and their host halos is built in.

\section{Conclusions}

In this paper, we have attempted to map out plausible ways to assemble
the $\gtrsim 10^{9}\Msol$ SMBHs that power the bright redshift
$z\approx 6$ quasars observed in the SDSS, without overproducing the
mass density in lower--mass ($\sim 10^{5-7}\Msol$) BHs.  We also
computed the event rates expected for {\it LISA} in each of the
successful models.

The physical effects governing SMBH assembly depend on the answers to
four basic questions: (1) how common are the initial BH seeds; (2) how
much mass in gas do they accrete, and therefore how much they
contribute individually to the final SMBH's mass; (3) how often do
they merge; and (4) what happens to SMBH binaries when they do merge?
Currently, we do not have empirical constraints to offer definitive
answers to any of these questions.  However, we are capable of
predicting the final outcome, starting with a set of assumptions for
the underlying physics.  Our trees-plus-orbits algorithm simulates the
formation history of SMBHs and the subsequent detection rate
expectations for {\it LISA} by isolating and prescribing answers to
the above four questions.  It is a powerful simulation tool, as it can
incorporate a detailed modeling of individual physical prescriptions
without a significant increase in the computational load, as long as
the prescriptions can be described by fitting formulae, tabulated in a
lookup table of reasonable size or summarized in a statistical manner.

Using this tool, we have surveyed a wide range of candidate assembly
models, and reported on common and distinguishing traits in the
resulting SMBH mass functions and the corresponding {\it LISA}
detection rates.  In particular, we have shown that SMBHs can form in
a manner consistent with other observational evidence either through
the rapid growth of rare, massive seeds, or through ultra--early
production of numerous Pop-III remnant seeds, provided these seeds
stop forming below a redshift $z_{\rm cut}\sim 20-30$.  We reach the
pessimistic conclusion that these scenarios do not produce {\it any}
detectable {\it LISA} events at $z>6$ (at least not in a few year's
operation).  An alternative model, in which we assume that the
extrapolation of the local $m-M$ relation holds at all redshifts
(e.g. due to internal feedback), on the other hand, can produce up to
$\sim 30$ {\it LISA} events per year, with a characteristic mass
spectrum.

Our major findings can be summarized more specifically as follows:

\begin{itemize}

\item SMBHs must be continuously surrounded by dense gas that was able
  to cool at the centers of DM halos.  Feeding holes with the
  low--density gas in DM halos whose gas was unable to cool does not
  allow for high enough accretion rates to explain the SDSS quasar
  BHs.

\item If embedded in dense gas nearly continuously, $\sim 100 \Msol$
  seed BHs can grow into the SDSS quasar BHs without super--Eddington
  accretion, but only if they form in minihalos at $z\gtrsim30$ and
  subsequently accrete $\gtrsim 60\%$ of the time.  However, these
  optimistic assumptions, required to explain the SDSS quasar BHs,
  overproduce the mass density in lower--mass (few$\times10^{5}\Msol
  \lsim M_{\rm bh}\lsim$ few$\times10^{7}\Msol$) BHs by a factor of
  $10^2-10^3$.  We find that two conditions need to be satisfied to
  alleviate this overprediction: the initial occupation fraction of
  seed BHs has to be low ($f_{\rm occ}\ltsim 10^{-2}$), and new seeds
  must stop forming, or the seeds must accrete at severely diminished
  rates or duty cycles, at $z\ltsim 20-30$.  We argued that models in
  which BH seeds stop forming at $z\sim 20$ are attractive because
  there are physical reasons for the seeds to stop forming below some
  redshift (such as metal pollution and/or radiative feedback that
  suppresses pop-III star formation), and because there is independent
  empirical evidence, from {\it WMAP} constraints on the reionization
  history, that star and/or BH formation in high--redshift minihalos
  was suppressed by a factor of $\gsim 10$ (Haiman \& Bryan 2006).

\item The simplest SMBH assembly scenarios, which have constant
  accretion rates, but in which BH seed formation stops abruptly at
  some redshift, and which meet constraints at both the high--mass and
  low--mass end of the $z=6$ SMBH mass function, predict negligibly
  low {\it LISA} event rates.  The reason for the low rates is as
  follows: in these models, the BHs that grow into the most massive,
  highest-redshift quasar-SMBHs accrete at the same (exponential) rate
  as all the other BHs, typically resulting in a vast overproduction
  of massive ($m\sim 10^{6}\Msol$) holes.  In order to offset this
  overproduction, seeds must be made very rare, and this diminishes
  the {\it LISA} rates.  It is difficult to envision a scenario for
  high ($\gtrsim 10$ per year per unit redshift) detection rates
  unless a vast number of SMBHs in the $10^{5-7}\Msol$ range lurk in
  the universe at all redshifts, which the current electromagnetic
  surveys have missed.

\item A different class of successful models, in which the SMBH masses
  are self--regulated by internal feedback, can evade this constraint,
  and produce {\it LISA} rates as high as $30$ yr$^{-1}$.  The key
  difference in these models that predict higher {\it LISA} rates is
  that the SMBH growth is driven by a large number of seed BHs and far
  lower gas accretion rates than those required in the
  constant-accretion models.  The majority of these events occur at
  $z\approx 6$ and in the low end ($10^3-10^4$~${\rm M_\odot}$) of {\it
  LISA}'s mass range for detection.
  Also, for these models we find the ejected BH mass
  density can exceed that of the galactic BH population at $z=6$.  Most
  ejected holes are expected to have masses similar to the seed mass,
  but an ejected BH can be as massive as $\sim10^{8}\Msol$ if large recoil
  velocities are allowed (e.g. if spins are not always aligned with
  the orbital angular momentum of the binary).

\end{itemize}

In addition to the above, our modeling reveals a number of interesting
aspects of SMBH assembly.  We find that in the successful models the
initial seeds are rare, and the most massive SMBHs grow primarily from
the few 'lucky' early seeds that avoided ejection due to kicks.  The
precise assumptions regarding the kick velocity distribution (such as
the assumed spin orientations or the resulting oscillation of the BH)
tend to have only a modest effect on the final results in these
models. This is because, at least in our simple prescription, BHs
either return quickly to the gas-rich nucleus or are left wandering in
the outer regions.  

Our results suggest that {\it LISA} will be capable of narrowing the
field of plausible SMBH assembly models from the raw event rate, even
without detailed measurements of the binary spins or mass ratios.  The
spin and mass ratio measurements will further constrain the evolution
of SMBH properties.  While the component prescriptions explored in
this paper are admittedly crude, exercises similar to the one
performed in our study will be crucial in understanding the limits and
possibilities offered by {\it LISA}, and ultimately to interpret the
detected {\it LISA} events.  The scarcity of empirical constraints on
the various pieces of physics that determines the SMBH growth leaves
us with a large range of ``plausible'' scenarios and free parameters
for SMBH assembly.

\acknowledgements

We thank Alessandra Buonanno for providing useful notes on the
gravitational recoil effect, and Greg Bryan and Marta Volonteri for
insightful conversations.  We also thank Kristen Menou for helpful
discussions and comments on the manuscript, and for providing
invaluable computational resources.  We also thank the anonymous
referee for a report that helped improve this manuscript.  This work
was supported by NASA through the ATFP grant NNX08AH35G.  ZH also
acknowledges support by the Pol\'anyi Program of the Hungarian
National Office of Technology.

\bibliographystyle{apj}

\begin{thebibliography}{99}
\bibitem[Abel et al. (2000)]{abn00} Abel, T., Bryan, G. L., \& Norman, M. L. 2000, ApJ, 540, 39
\bibitem[Abel et al. (2002)]{abn02} Abel, T., Bryan, G. L., \& Norman, M. L. 2002, Science 295, 5552, 93
\bibitem[]{awa08} Alvarez, M., Wise, J. H., \& Abel, T. 2008, ApJL, submitted, preprint arXiv:0811.0820
\bibitem[Arun et al. (2008)]{a08} Arun, K.G., et al. 2008, in Classical and Quantum Gravity, special issue with contributions to the proceedings of 7th LISA Symposium, in press, gr-qc/0811.1011v1
\bibitem[]{b06} Baker, J.G., Centrella, J.; Choi, D.; Koppitz, M., \& van Meter, J. 2006, Phys. Rev. Lett. 96, 111102
\bibitem[]{b08} Baker, J.G., Boggs, W.D., Centrella, J., Kelly, B.J., McWilliams, S.T., Miller, M.C., van Meter, J.R. 2008 ApJ, 682, L29
\bibitem[]{b04} Barkana, R. 2004, MNRAS 347, 57
\bibitem[]{bl01} Barkana, R., \& Loeb, A. 2001, Physics Reports, 349, 125
\bibitem[]{bvr06} Begelman, M., Volonteri, M., \& Rees, M. J. 2006, MNRAS, 370, 289
\bibitem[]{bmsb06} Berczik, P., Merritt, D., Spurzem, R., Bischof, H. 2006, ApJ 642, L21
\bibitem[]{bv08}Berti, E., \& Volonteri, M. 2008, ApJ, 684, 822B
\bibitem[]{bt88} Binney, J., \& Tremaine, S. 1988, {\it Galactic Dynamics}, Princeton University Press
\bibitem[]{bl08} Blecha, L. \& Loeb, A. 2008, MNRAS, 390, 1311
\bibitem[]{brm07} Bogdanovic, T., Reynolds, C.S., Miller, M.C. 2007, ApJ 661L, 147B
\bibitem[]{bss07} Bonning, E. W., Shields, G. A., \& Salviander, S. 2007, ApJ 666, 13
\bibitem[]{bmq06} Boylan-Kolchin, M., Ma, C., Quataert, E. 2006, MNRAS 369, 1081
\bibitem[]{bmq08} Boylan-Kolchin, M., Ma, C., Quataert, E. 2008, MNRAS 383, 93
\bibitem[]{bsf04} Bromley, J.M., Somerville, R.S., \& Fabian, A.C. 2004, MNRAS 350, 456
\bibitem[]{bcl02} Bromm, V., Coppi, P. S., Larson, R. B. 2002, ApJ, 564, 23
\bibitem[Bromm \& Loeb(2003)]{bl03} Bromm, V., \& Loeb, A. 2003, ApJ, 596, 34
\bibitem[]{c06} Campanelli, M., Lousto. C., Marronetti, P., \& Zlochower, Y.  2006, Phys. Rev. Lett. 96, 111101
\bibitem[]{c00}Cole, S., Lacey, C. G., Baugh, C. M., Frenk, C. S. 2000, MNRAS, 319, 168
\bibitem[]{d04} Dijkstra, M., Haiman, Z., Rees, M.J., Weinberg, D.H. 2004, ApJ 601, 2, 666D
\bibitem[]{eh99} Eisenstein, D., \& Hu, W. 1999, ApJ, 511, 5
\bibitem[]{e04} Escala, A., Larson, R., Coppi, P., \& Mardones, D. 2004, ApJ 507, 765
\bibitem[]{f01} Fan, X., et al. 2001, ApJ 122, 2833
\bibitem[]{f03} Fan, X., et al. 2003, AJ 125, 1649
\bibitem[Fan(2006)]{fan06} Fan, X. 2006, New Astronomy Reviews, 50, 665
\bibitem[]{fhh04} Favata, M., Hughes, S.A., \& Holz, D.E. 2004, ApJ, 607, L5
\bibitem[Haehnelt(2003)]{h03} Haehnelt, M. 2003, Class. Quantum Grav., 20, 31
\bibitem[]{fm00} Ferrarese, L. \& Merritt, D. 2000, ApJ, 539, L9
\bibitem[]{f02} Ferrarese, L. 2002, ApJ, 578, 90
\bibitem[]{g00} Gebhardt, K. et al. 2000, ApJ, 539, L13
\bibitem[]{g08} Guedes, J. Diemand, J. Zemp, M., Kuhlen M., Madau P., Mayer L., \& Stadel, J. 2008, in Proceedings of the conference Galactic \& Stellar Dynamics In the Era of High Resolution Surveys, 2008, AN, 329, 1004G
\bibitem[]{h97a}Haiman, Z., Rees, M.J., \& Loeb, A. 1997, ApJ, 476, 458H
\bibitem[]{h97b}Haiman, Z., Rees, M.J., \& Loeb, A. 1997, ApJ, 484, 985H
\bibitem[]{h04} Haiman, Z. 2004, ApJ 613, 36
\bibitem[]{hb06} Haiman, Z. \& Bryan, G. L. 2006, ApJ 650, 7
\bibitem[Haiman, Abel \& Rees(2000)]{HAR2000} Haiman, Z., Abel, T., \& Rees, M. J. 2000, ApJ, 534, 11
\bibitem[]{hco04} Haiman, Z., Ciotti, L, \& Ostriker, J.P. 2004, ApJ 696, 763H
\bibitem[]{hjb07} Haiman, Z., Jimenez, R., \& Bernardi, M. 2007, ApJ, 658, 721H
\bibitem[]{hl01} Haiman, Z., \& Loeb, A. 2001, ApJ 552, 459
\bibitem[]{htl96} Haiman, Z., Thoul, A., Loeb, A. 1996 ApJ 464, 523
\bibitem[Heger et al. (2003)]{hetal03} Heger, A., et al. 2003, ApJ, 591, 288
\bibitem[]{h02} Hughes, S.A. 2002, MNRAS 331, 805
\bibitem[Islam, Taylor \& Silk (2004)]{its04} Islam, R. R., Taylor, J. E., \& Silk, J. 2004, \mnras, 354, 629
\bibitem[]{kkh05} Keeton, C., Kuhlen, M., \& Haiman, Z. 2005, ApJ 621, 559
\bibitem[]{k95} Kidder, L. E. 1995, Phys. Rev. D, 52, 821
\bibitem[]{k08} Komatsu, E. et al., 2008, ApJS, 180, 330K
\bibitem[]{kzl08} Komossa, S., Zhou, H., \& Lu, H. 2008 ApJ, 678, L81
\bibitem[]{kbd04} Koushiappas, S.M., Bullock, J.S., Dekel, A., 2004 MNRAS 354, 292
\bibitem[Koushiappas \& Zentner(2006)]{kz06} Koushiappas, S. M., \& Zentner, A. R. 2006, ApJ, 639, 7
\bibitem[]{lc92} Lacey, C., \& Cole, S. 1993, MNRAS 262, 627
\bibitem[Lang \& Hughes(2006)]{lh06} Lang, R. \& Hughes, S. A. 2006, \prd, 74, 122001; erratum 2007, \prd, 75, 089902
\bibitem[Lippai, Frei \& Haiman(2008)]{lfh08} Lippai, Z., Frei, Z., \& Haiman, Z. 2008b, ApJL, 676L, 5L
\bibitem[]{ln06} Lodato, G., \& Natarajan, P. 2006, MNRAS, 371, 1813
\bibitem[]{mba01} Machacek, M., Bryan, G., \& Abel, T. 2001, ApJ 548, 509
\bibitem[]{mq04} Madau, P., \& Quataert, E. 2004, ApJ 606, L17
\bibitem[]{m07} Mayer, L., Kazantzidis, S., Madau, P., Colpi, M., Quinn, T., \& Wadsley, J. 2007, Sci 316.1874M
\bibitem[Menou(2003)]{m03} Menou, K. 2003, Class. Quantum Grav., 20, 37
\bibitem[]{mh04} Menou, K., \& Haiman, Z. 2004, ApJ, 615, 130
\bibitem[Menou, Haiman \& Narayanan(2001)]{mhn01} Menou, K., Haiman, Z., \& Narayanan, V. K 2001, \apj, 558, 535
\bibitem[]{mbh06}Mesinger, A., Bryan, G.L., \& Haiman, Z. 2006, ApJ, 648, 835M
\bibitem[Micic et al.(2007)]{mic07} Micic, M., Holley-Bockelmann, K., Sigurdsson, S. \& Abel, T. 2007, MNRAS, 380, 1533M
\bibitem[]{mm03} Milosavljevic, M., \& Merritt, D. 2003, ApJ 596
\bibitem[]{nfw97} Navarro, J., Frenk, C.,\& White, S. 1997, ApJ 490, 493N
\bibitem[]{oh02} Oh, S. P., \& Haiman, Z. 2002, ApJ, 569, 558
\bibitem[]{osh08} Omukai, K., Schneider, R., \& Haiman, Z. 2008, ApJ, 686, 801
\bibitem[]{on07} O'Shea, B. \& Norman, M. 2007, ApJ 654, 66
\bibitem[]{o99}Ostriker, E.C. 1999, ApJ, 513, 252O
\bibitem[]{} Press, W.H., \& Schechter, P. 1974, ApJ, 187, 425P
\bibitem[]{p05} Pretorius, F. 2005, Phys. Rev. Lett. 95, 121101
\bibitem[]{r04} Richards, G. T., Strauss, M. A., Pindor, B., Haiman, Z., Fan, X., Eisenstein, D., et al. 2004, AJ, 127, 1305
\bibitem[Schnittman \& Buonanno(2007)]{sb07} Schnittman, J., \& Buonanno, A. 2007, ApJ, 662, 63
\bibitem[Schnittman et al.(2008)]{setal08} Schnittman, J. D., Buonanno, A., van Meter, J. R., Baker, J. G., Boggs, W. D., Centrella, J., Kelly, B. J., \& McWilliams, S. T. 2008, Phys. Rev. D, 77, 4031
\bibitem[Sesana et al.(2004)]{ses04} Sesana, A., Haardt, F., Madau, P. \& Volonteri, M. 2004, \apj, 611, 623
\bibitem[Sesana et al.(2005)]{ses05} Sesana, A., Haardt, F., Madau, P. \& Volonteri, M. 2005, \apj, 623, 23
\bibitem[Sesana et al.(2007)]{ses07} Sesana, A., Volonteri, M. \& Haardt, F. 2007, MNRAS, 377, 1711S
\bibitem[]{scmfw08} Shankar, F., Crocce, M., Miralda-Escud\'e, J., Fosalba, P., \& Weinberg, D. H. 2008b, ApJ, submitted, preprint arXiv:0810.4919
\bibitem[]{swm09a} Shankar, F., Weinberg, D. H., \& Miralda-Escud\'e, J. 2009a, ApJ, 690, 20S
\bibitem[]{sbh09b} Shankar, F., Bernardi., M., Haiman, Z. 2009b, ApJ, 694, 867S
\bibitem[]{s05} Shapiro, S. L. 2005, ApJ, 620, 59
\bibitem[]{sir99} Shapiro, P. R., Ilyev, I., \& Raga, A. C. 1999, MNRAS, 307, 203
\bibitem[]{s07} Shen, Y. et al. 2007, AJ, 133, 2222
\bibitem[]{sk05} Somerville, R., \& Kolatt, T. 2005, MNRAS 305, 1
\bibitem[Spaans \& Silk(2006)]{ss06} Spaans, M., \& Silk, J. 2006, ApJ, 652, 902
\bibitem[Turner (1991)]{t91}Turner, E.L. 1991AJ, 101, 5T
\bibitem[]{t97} Tegmark, M. et al. 1997, ApJ 474, 1
\bibitem[Vecchio (2004)]{vec04} Vecchio, A. 2004, \prd, 70, 042001
\bibitem[]{vhm03} Volonteri, M., Haardt, F., \& Madau, P. 2003, ApJ 582, 559
\bibitem[]{vr05} Volonteri, M., \& Rees, M. J. 2005, ApJ, 633, 624
\bibitem[]{vr06} Volonteri, M., \& Rees, M. J. 2006, ApJ, 650, 669
\bibitem[Willot et al. (2003)]{}Willott, C.J., McLure, R.J., \& Jarvis, M. 2003, ApJ 587L, 15W
\bibitem[Willot et al. (2006)]{}Willott, C.J., et al. 2006, NOAO Proposal 383W, ID \#2006B-0383
\bibitem[Whalen, Abel \& Norman(2004)]{wan04} Whalen, D., Abel, T. \& Norman, M. L. 2004, ApJ, 610, 14
\bibitem[Wyithe \& Loeb(2003)]{wl03} Wyithe, J. S., \& Loeb A. 2003, \apj, 590, 691
\bibitem[]{ym04} Yoo, J., \& Miralda-Escude, J. 2004, ApJ 614, L25
\bibitem[]{y03} Yoshida, N., Abel, T., Hernquist, L., \& Sugiyama, N. 2003, ApJ, 592, 645
\bibitem[]{yoh08} Yoshida, N., Omukai, K., Hernquist, L. 2008, Science, 321, 669Y
\bibitem[]{yt02} Yu, Q., \& Tremaine, S. 2002, MNRAS 335, 965
\bibitem[]{z08} Zhang, J., Fakhouri, O, \& Ma, C.-P. 2008, MNRAS, 389, 1521
\end{thebibliography}

\clearpage

\begin{table}[ht]
\caption{Masses and Quantities of Simulated DM Halos.}
\centering
\begin{tabular} {c c  c c c}
\hline\hline
$M_{\rm lo}<M<M_{\rm hi}$ & $\log_{10}N_{\rm bin}$
&$ \log_{10}\langle M_{\rm bin}\rangle$ &  $N_{\rm sim}$ & $W_{\rm bin}$\\
[0.5 ex]
\hline
$8.0 <\log_{10}M<8.5$ & $12.33$ & $8.22$ & $50000$ & $4.28\times 10^{7}$ \\
$8.5 <\log_{10}M<9.0$ & $11.78$ & $8.72$ & $27000$ & $2.23\times 10^{7}$\\
$9.0 <\log_{10}M<9.5$ & $11.20$ & $9.22$ & $15000$ & $1.06\times 10^{7}$\\
$9.5 <\log_{10}M<10.0$ & $10.57$ & $9.71$ &$9000$ & $4.13\times 10^{6}$\\
$10.0 <\log_{10}M<10.5$ & $9.87$ & $10.2$ & $5000$ & $1.48\times 10^{6}$\\
$10.5 <\log_{10}M<11.0$ & $9.08$ & $10.7$ &$2700$& $ 4.45\times 10^{5}$\\
$11.0 <\log_{10}M<11.5$ & $8.14$ & $11.2$ &$1500$ & $9.20\times 10^{4}$\\
$11.5 <\log_{10}M<12.0$ & $6.98$ & $11.7$ &$900$ & $1.06\times 10^{4}$\\
$12.00 <\log_{10}M<12.5$ & $5.50$ & $12.1$ &$500$ & 632\\
$12.50 <\log_{10}M<12.85$ & $3.48$ & $12.6$ &$303$ & 10.\\
$12.85 <\log_{10}M<\infty$ & $1.60$ & $12.9$ & $40$ & 1.0\\
[1ex]
\hline
\end{tabular}
\end{table}
For each mass bin, BH assembly
was simulated by creating a merger tree for $N_{\rm sim}$ Monte-Carlo halos, and the results
were multiplied by the weighting factors $W_{\rm bin}$
to represent the Press-Schecter mass function for DM
halos at $z=6$.  $N_{\rm bin}=\int_{M_{\rm lo}}^{M_{\rm hi}}dN/dM\; dM$
is the expected Press-Schechter number of halos in each bin, and
$\langle M_{\rm bin} \rangle$ is the number-weighted mean halo mass in each bin.\\

\begin{table}[ht]
\caption{Properties of four successful (A-D) and two failed (X and Y)
models for SMBH growth} \centering
\begin{tabular} {c c  c c c c c c}
\hline\hline
Model &
$m_{\rm seed}$ &
$T_{\rm seed}$ &
$ f_{\rm seed}$ &
$f_{\rm duty}$ &
spin &
$z_{\rm cut}$ &
$\rho_{\rm SMBH, 5+} (z=6)$
\\
[0.5 ex]
\hline
A & $200 \Msol$ & $1200\K$ & $10^{-4}$ & $1$ & aligned & 25
&  $3.4\times 10^{4}\Msol \Mpc^{-3}$\\
B & $100 \Msol$ & $1200\K$ & $10^{-2}$ & $0.95$ & aligned & 28 &
$5.1\times 10^{4}\Msol \Mpc^{-3}$\\
C & $10^{5} \Msol$ & $1.5\times 10^{4} \K$ & $10^{-4}$ & $0.6$ & random & 13 & $6.2\times 10^{4}\Msol \Mpc^{-3}$\\
D & $2\times 10^{5} \Msol$ & $1.5\times 10^{4}  \K$ & $10^{-2}$ & $0.55$ & aligned & 18 &$ 7.0\times 10^{4}\Msol \Mpc^{-3}$\\
X & $100 \Msol$ & $1200  \K$ & $1$ & $0.8$ & random & 0 &$ 2.9\times 10^{8}\Msol \Mpc^{-3}$\\
Y & $10^{5} \Msol$ & $1.5\times 10^{4}  \K$ & $10^{-3}$ & $0.6$ & aligned & 0 &$ 1.1\times 10^{6}\Msol \Mpc^{-3}$\\
[1ex]
\hline
\end{tabular}
\end{table} 
The above shows parameters for four models that (1) have constant
accretion rates of $f_{\rm duty}$ times the Eddington rate; (2)
produce by $z=6$ SMBHs massive enough to power the SDSS quasars; and
(3) do not overproduce the overall SMBH population.  In Models A and B
the seed BHs are Pop III remnants, and in Models C and D the seeds are
formed through direct--collapse in more massive halos.  Models X and Y
are unsuccessful models that barely produce the $m\gtrsim 10^{9}\Msol$
SMBHs by $z=6$ but also far overproduce the lower-mass SMBHs.

\begin{table}[ht]
\caption{The total seed, total ejected, and total retained BH masses}
\centering
\begin{tabular} {c  c c c c}
\hline\hline
Model 
&$m_{\rm seed,\; tot}$ & $m_{\rm GN}(z=6)$ 
& $m_{\rm ejected,\; tot}$
& $m_{\rm BH,\; tot}/m_{\rm seed,\; tot}$
\\
[0.5 ex]
\hline
A & $9.3$ & $16.0$ & $6.2$ & $6.7$  \\
B & $10.1$ & $16.1$ & $7.7$ & $6.0$  \\
C & $12.7$ & $16.2$ & $6.4$ & $3.4$  \\
D & $13.2$ & $16.3$ & $6.2$ & $3.1$  \\
X & $16.0$ & $20.0$ & $19.1$ & $4.0$  \\
Y & $14.6$ & $17.5$ & $12.7$ & $2.9$  \\
$m$-$\sigma$, $z_{\rm cut}=0$, $f_{\rm seed}=1$, aligned & $15.7$ & $15.7$ & $16.0$ & $0.48$  \\
$m$-$\sigma$, $z_{\rm cut}=0$, $f_{\rm seed}=1$, random & $15.7$ & $15.7$ & $16.3$ & $0.70$  \\
$m$-$\sigma$, $z_{\rm cut}=0$, $f_{\rm seed}=10^{-2}$, aligned & $13.7$ & $15.6$ & $12.9$ & $1.9$  \\
$m$-$\sigma$, $z_{\rm cut}=18$, $f_{\rm seed}=1$, aligned & $14.4$ & $15.7$ & $14.7$ & $1.34$  \\
[1ex]
\hline
\end{tabular}
\end{table}
The decomposition of the final SMBH mass into the contributions from
the initial stellar seed BHs and subsequent gas accretion. Masses are
in $\log_{10}$, and in units of $\Msol$.  The columns, starting from
the second and from left to right, show: the total initial seed mass;
the total SMBH mass retained in nuclei at $z=6$ ; the total ejected BH
mass, and $\log_{10}$ of the ratio of the total (i.e. the sum of the
ejected and nuclear) $z=6$ SMBH mass to the initial seed mass.  The
last ratio is a measure of the total growth due to gas accretion.  The
first four rows (A-D) show values for the four successful models. Also
shown for comparison are values from two of the unsuccessful,
constant-accretion models (X and Y) that overproduce the total BH mass
function.  The model parameters for Models A, B, C, D, X and Y can be
found in Table 2.  The models where the $m$-$\sigma$ relation is
enforced by hand can grow primarily through mergers, with gas
accretion adding as little as a factor of a few to the total SMBH mass
at $z=6$.  If $f_{\rm seed}$ is sufficiently high, they also eject a
total mass in low-mass BHs that is comparable to the retained nuclear
population (most of the ejected holes have a mass near the seed mass).
\\

\clearpage

\begin{figure}
\centerline{\hbox{
\plotone{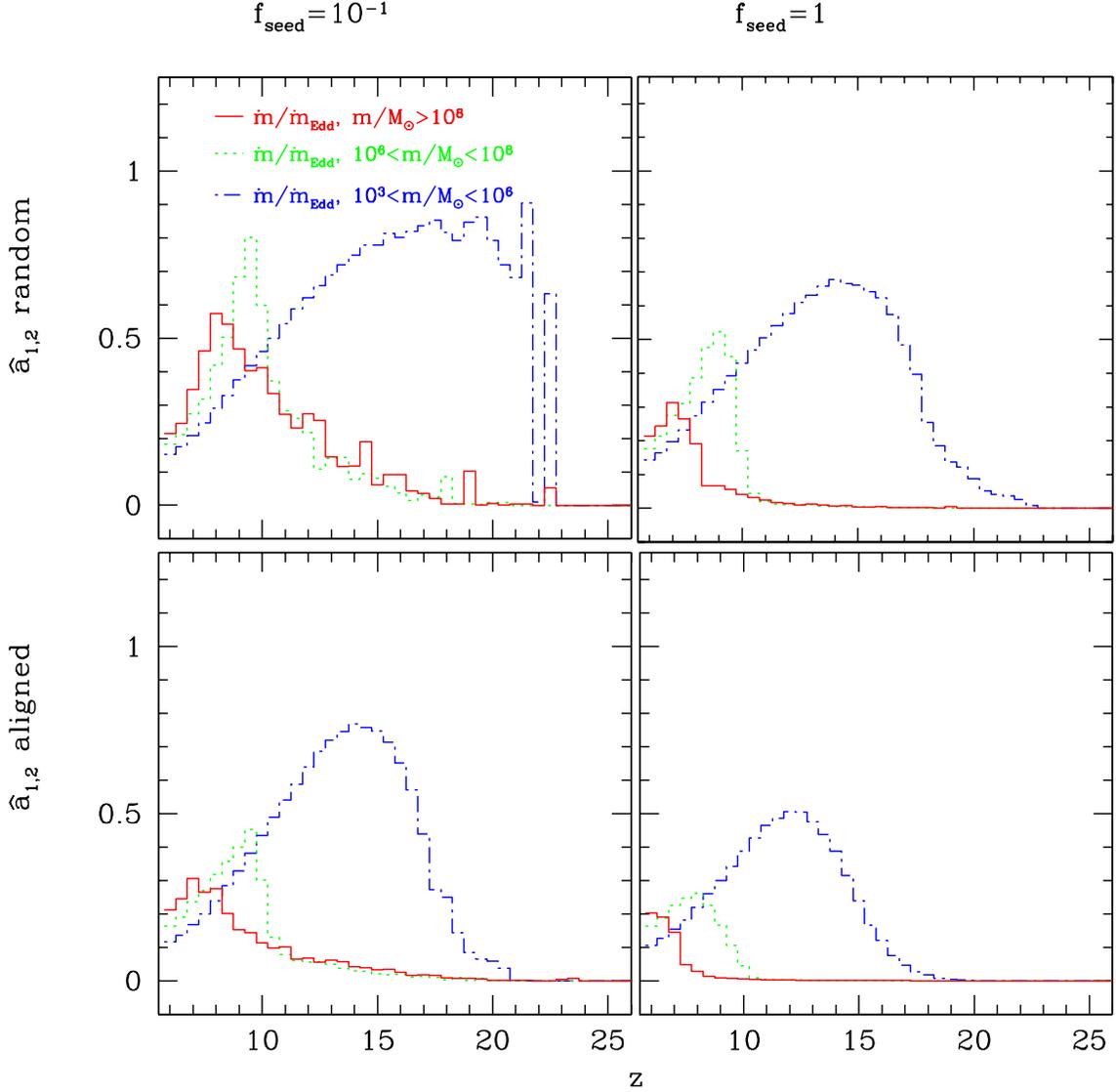}
}}
\caption{The accretion rate (thick lines) in Eddington limits in
  models that match the $m$-$\sigma$ relation (Equation \ref{mM}) at all
  redshifts, for different ranges of BH masses.  We also plot the
  merger rates (thin lines) in units of mergers per halo for
  different halo mass bins corresponding to $M_{1}+M_{2}=10^{4}m$.
  Note that
  masses are defined instantaneously at each redshift interval,
  rather than tracking the histories of the $z=6$ holes and halos.
  }
  \label{fig:accms}
  \end{figure}

\end{document}